\pdfoutput=1
\documentclass{optica-article}

\journal{opticajournal} %

\articletype{Research Article}

\usepackage{lineno}

\usepackage{amsmath}
\usepackage{amsthm}
\usepackage{physics}
\usepackage{booktabs}
\usepackage{placeins}

\newcommand{\dvec}{\vb*{d}}
\newcommand{\Lam}{\Lambda}
\newcommand{\Lmax}{\Lam_{\text{max}}}
\newcommand{\nrm}{\vu*{\nu}}
\newcommand{\q}[1]{q\qty(#1)}
\newcommand{\Phinear}{\Phi_{\text{near}}}
\newcommand{\Phifar}{\Phi_{\text{far}}}
\newcommand{\Znear}{Z_{\text{near}}}
\newcommand{\Zfar}{Z_{\text{far}}}
\newcommand{\Wnear}{W_{\text{near}}}
\newcommand{\Wfar}{W_{\text{far}}}

\graphicspath{{figures/}}
\newcommand{\paperfig}[2][\linewidth]{\IfFileExists{figures/#2.pdf}{\includegraphics[width=#1]{#2}}%
  {\fbox{\parbox[c][6em][c]{#1}{\centering\footnotesize figure file \texttt{\detokenize{#2}.pdf} not built yet}}}}

\begin{document}

\title{A complete forward projection algorithm through cylindrical and spherical refractive windows}

\author{Shijie Zhong\authormark{1}, Shiyong Tan\authormark{1}, Xiaosi Zhang\authormark{1},  Yanying Li\authormark{2}, Rui Ni\authormark{1,*}}

\address{\authormark{1}Johns Hopkins University, Baltimore, Maryland 21218, USA}
\address{\authormark{2}Garrison Forest School, Owings Mills, Maryland 21117, USA}

\email{\authormark{*}rui.ni@jhu.edu}

\begin{abstract*}
Cylindrical or spherical windows often appear in experiments and real applications. Measuring through those windows requires projecting 3D points to pixels, which is different to compute due to refraction. A point may have multiple projected images but cannot be efficiently determine with the existing projection algorithms. Using the ray's conserved momenta, we reduce this projection to closed-form equations with a concave structure. As a result, we can successfully determine whether a 3D point has a single projection or multiple and compute those locations. In simulations, with one single search, we can recover images that iterative methods miss even with many restarts. A wand calibration based on it can reach the precision limit set by image noise, enabling accurate 3D particle tracking in a swirling flow through cylinder windows. This algorithm shows how to predict visibility and multiple images through curved windows, shedding light for 3D pipe flow measurements, accurate underwater reconstruction, and many real applications.
\end{abstract*}

\section{Introduction}
\label{sec:intro}
Many optical measurements need to look through a curved transparent wall or interface. 
Flow velocimetry is performed inside pipes~\cite{lowe1992refraction,bauerlein2025pulsatile,liu2013investigation}, cylindrical tanks~\cite{belden2013calibration,paolillo2022perspective}, Taylor--Couette cells~\cite{huisman2012applying}, and even in droplets~\cite{kang2004quantitative}. 
People also track fish in cylindrical aquaculture tanks to study schooling~\cite{abe2020multiple}.
In combustion, flames are reconstructed inside the glass cylinders~\cite{liu2019computed}, and optical engines~\cite{zha2016novel,gao2023volumetric}.
Sometimes the window is made curved to sustain high pressure: underwater cameras usually sit in pressure housings or dome port~\cite{kunz2008hemispherical,nocerino2016underwater}, which has spherical windows~\cite{she2022refractive,menna2016geometric,nocerino2016underwater}.

\begin{figure}[htbp]
\centering
\paperfig[\linewidth]{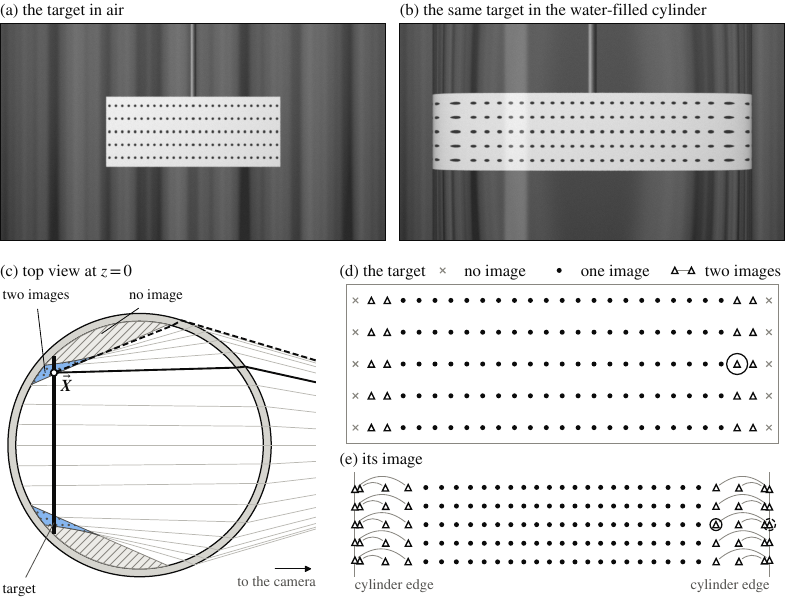}
\caption{Simulated projection of a $27\times5$ dot target through the reference cylindrical window: air--acrylic--water with indices $(1,1.49,1.33)$ and outer/inner radii $(100,94)$\,mm. The pinhole and target plane lie 350\,mm and 65\,mm from the axis on opposite sides; dots are spaced 5\,mm laterally and 10\,mm axially. (a)--(b)~Simulated camera images of the target, held by a rod, in air and inside the water-filled cylinder, rendered by tracing each pixel's ray with vector Snell's law. (c) Top view at $z=0$: hatched and dotted regions are reached by zero and two rays, respectively; solid and dashed rays connect the marked dot $\va*{X}$ to the pinhole. (d) Target dots classified by image count: 10 have no image, 105 have one, and 20 have two. (e) Computed image in normalised coordinates: arcs link paired images, vertical lines mark the cylinder edges, and circles identify the two images of $\va*{X}$. Every computed ray was confirmed by vector-Snell tracing.}
\label{fig:teaser}
\end{figure}

Refraction through a curved wall bends the lines of sight, which distorts the images of objects inside a cylinder, and some can be hidden or seen more than once~\cite{lowe1992refraction,paolillo2022perspective}. 
As shown in Fig.~\ref{fig:teaser}(a) and (b), the simulated dot plate image is stretched horizontally after putting into a water-filled acrylic cylinder. The outmose columns disappear through the refraction, while the next columns have two projected images.
To avoid such refraction effects, matching refraction index of the fluid is often used~\cite{paolillo2022perspective,bauerlein2025pulsatile}. However, constrains the fluid is practically impossible in gases~\cite{paolillo2022perspective}, or in a large scale cylindrical fish tank~\cite{abe2020multiple}.
Therefore, we need to develop a complete forward projection algorithms of any 3-D point through the curved wall. Only with that, camera calibration~\cite{zhang2000flexible,hartley2004multiple}, and the stereo matching between cameras~\cite{schanz2016shake,tan2020openlpt} are possible.

Forward projection through flat interfaces is well understood, where all interfaces share one normal and the projection can be reduced to closed-form equations.
For one window, Agrawal et al. expressed forward projection as a fourth- or twelfth-degree polynomial equation~\cite{agrawal2012theory}.
Later, Gu et al. reduced multilayer projection to one monotone equation~\cite{gu2023refraction}, by using the property that the shared normal also keeps the ray's tangential momentum constant across layers.
In addition to these closed-form algorithms, virtual-camera models offer another representation of flat window imaging~\cite{treibitz2012flat,kawahara2013pixelwise}.

However, curved interfaces is much harder because their normals vary along the interfaces and the symmetric properties of flat-window cannot transfer directly.
For outward-looking underwater dome cameras, Kunz and Singh modeled projection numerically~\cite{kunz2008hemispherical}, while She et al. derived a $6^{\text{th}}$ degree equation for a thin dome and used iterative projection for a thick dome~\cite{she2022refractive}.
In planar-wall and cylindrical-tank calibration, Belden alternated forward and backward ray traces~\cite{belden2013calibration}. Paolillo and Astarita used iterative dewarping for complex geometries, including a cylindrical tank~\cite{paolillo2022perspective}.
A polynomial equation can yield several roots, some of which may not describe rays that can pass through the interfaces. By contrast, an iteration started from one guess returns at most one candidate ray.

The differences in output matter when a point is reached by no ray or by several. Paolillo and Astarita showed numerically that both cases occur for a water-filled PMMA cylinder viewed from air~\cite{paolillo2022perspective}. A search started from one guess may find only one of several rays, while a failed search cannot guarantee that no ray exists.
To the best of our knowledge, prior work has neither characterised visibility and image count nor provided a search for every image through the inward-looking multilayer cylinders and spheres considered here.

The illustration of the visible regions, single and multi image regions of a dot target is shown in Fig.~\ref{fig:teaser}.
Fig.~\ref{fig:teaser}(c) shows a top view of the cylindrical window. The hatched regions cannot be reached by a ray, while the dotted blue regions can be reached by two rays. 
To demostrate the effects of different regions, we simulate a 135-dot target as shown in Fig.~\ref{fig:teaser}(d). The cross symbol means no image, the filled circle means one image, and the triangle means two images.
In the projected pattern shown in Fig.~\ref{fig:teaser}(e), the outermost columns therefore disappear, while the doubly imaged dots produce additional images near the cylinder edges.
A solver that returns only one ray misses these additional images, while failed convergence alone cannot guarantee the invisible dots.
This distinction matters when matching detections to 3D points or selecting observations for calibration.

In this paper, we consider cameras outside coaxial cylindrical or concentric spherical windows, looking at points inside.
A ray keeps two quantities as it crosses the windows: its angular momentum about the axis and its momentum along the axis, and only one angular momentum is needed for spherical windows due to symmetry~\cite{born1999principles}.
These conserved quantities are classical and also used for optical fibres and seismic rays~\cite{snyder1983optical,aki2002quantitative}.
With them, a ray can be described by two numbers instead of guessing a full path by iterations. 
Based on them, closed form equations can be derived by constraints of how far it turns around the axis and advances along the axis~\cite{buland1983computation}.
The main contributions of this paper are:
\begin{enumerate}
\item \textbf{Complete solution to forward projection.} We derive closed-form ray equations for multilayer cylinders and spheres, and develop a forward projection algorithm. This algorithm can return the complete list of a point's images, including an empty if it is invisible. It is implemented in C++ for OpenLPT library~\cite{tan2020openlpt,tan20233d}, and the code is available at \url{https://github.com/clockj/pyOpenLPT-refraction}.
\item \textbf{visibility region maps.} Based on the forward projection algorithm, we can find the region maps for empty, single and multi images like Fig.~\ref{fig:teaser}(c). So people can focus on the single image region for calibration.
\item \textbf{Calibration and tracking methods.} A wand calibration method is developed without the need of dot plates, enabling the calibration in complicated experimental setup. We also show the ability for Lagrangian particle tracking through a cylinder wall for swirling flow.
\end{enumerate}

The structure of the paper is listed as follows. Sections~\ref{sec:geometry} and~\ref{sec:flat} explain the geometry and flat-window reference. Then Sections~\ref{sec:equations} and~\ref{sec:near} shows the derivation of the curved-window equations and near-branch solver. Section~\ref{sec:visibility} analyzes visibility and image count, followed by the summary of algorithm in Section~\ref{sec:algorithm}. The special case of spherical windows is discussed in Section~\ref{sec:sphere}. Finally, Section~\ref{sec:validation} presents a wand calibration built on the projection and validates both through synthetic tests and one real experiment, and Section~\ref{sec:conclusion} shows the conclusion.

\section{Geometry and conserved quantities}
\label{sec:geometry}
\subsection{Setup and notation}
\label{sec:setup}

\begin{figure}[htbp]
\centering
\begin{tabular}{@{}c@{}}
\paperfig[0.68\linewidth]{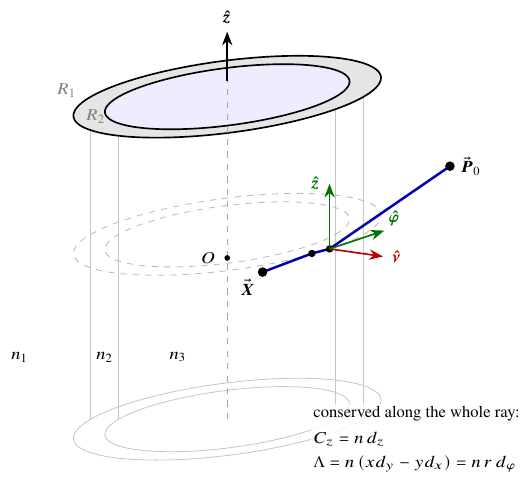} \\ (a) \\[2mm]
\paperfig[0.62\linewidth]{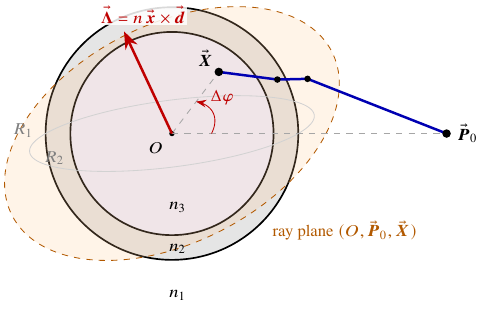} \\ (b)
\end{tabular}
\caption{Ray geometry for three media separated by interfaces $R_1>R_2$: medium $1$ contains the pinhole $\va*{P}_0$, medium $2$ is the wall, and medium $3$ contains the object point $\va*{X}$. (a) Coaxial cylinders. The interface normal $\vu*{\nu}$ is radial, while $\vu*{\varphi}$ and $\vu*{z}$ are tangential directions. The skew ray conserves $C_z=nd_z$ and $\Lam=nr d_\varphi$, where $d_\varphi=\dvec\vdot\vu*{\varphi}$. (b) Concentric spheres. The conserved vector $\va*{\Lambda}=n\,\va*{x}\cross\dvec$ is normal to the ray plane $(O,\va*{P}_0,\va*{X})$; $\Delta\varphi$ is the central angle between $O\va*{P}_0$ and $O\va*{X}$.}
\label{fig:geometry}
\end{figure}

Suppose we want to measure the flow inside a transparent pipe or to calibrate a camera behind a dome window. As shown in Fig.~\ref{fig:geometry}, a point in the field of view, $\va*{X}$, can be projected through $N-1$ coaxial circular-cylindrical or concentric spherical interfaces onto the camera focal plane through the pinhole point, $\va*{P}_0$.

Here, medium $1$ contains the pinhole $\va*{P}_0$, medium $N$ contains the object point $\va*{X}$, and medium $k$ with refractive index $n_k$ has range $R_k\le r\le R_{k-1}$ ($r$ is the radial distance to the central axis of cylinder or center of sphere). 
$R_1>R_2>\dots>R_{N-1}$ are the interface radii, and $R_0\equiv r_P$ and $R_N\equiv r_X$ are the radial coordinates of $\va*{P}_0$ and $\va*{X}$ (Fig.~\ref{fig:index} in Appendix~\ref{app:index}). Table~\ref{tab:symbols} collects the recurring notation; quantities used only in individual steps are defined where they arise.
Unless stated otherwise, numerical examples use a \emph{reference window}: a water-filled acrylic cylinder viewed from air, with interface radii $(R_1,R_2)=(100,94)$\,mm, refractive indices $(n_1,n_2,n_3)=(1,1.49,1.33)$, and pinhole distance $r_P=350$\,mm from the axis. The object radius $r_X$ varies between examples.

\begin{table}[!htbp]
\caption{Principal notation. Unless stated, ray quantities refer to the cylindrical geometry.}
\label{tab:symbols}
\centering
\small
\begin{tabular}{@{}p{0.28\linewidth}p{0.67\linewidth}@{}}
\toprule
Symbol & Meaning \\
\midrule
$\va*{P}_0$, $\va*{X}$ & pinhole and object point \\
$N$, $k$, $n_k$ & number of media, medium index, and refractive index of medium $k$ \\
$r$, $\varphi$, $z$, $R_k$ & radial, azimuthal, and axial coordinates; radius bounding medium $k$ ($R_k\le r\le R_{k-1}$) \\
$R_0=r_P$, $R_N=r_X$ & radial coordinates of the pinhole and object point \\
$\Delta\varphi$, $\Delta z=z_X-z_P$ & azimuthal separation and axial offset from $\va*{P}_0$ to $\va*{X}$ \\
$\dvec$, $d_\varphi$, $d_z$ & unit ray direction and its azimuthal and axial components \\
$\Lam=nr d_\varphi$, $C_z=nd_z$ & conserved angular and axial momenta for a cylinder \\
$\nu_k$, $b_k=\Lam/\nu_k$ & effective transverse index $\sqrt{n_k^2-C_z^2}$ and impact parameter in medium $k$ \\
$\Delta\varphi_k$, $s_k$, $\Delta z_k$ & azimuthal sweep, transverse chord length, and axial travel in medium $k$ \\
$\Lmax$, $k^\ast$ & limiting angular momentum and the medium setting that limit \\
$\Phinear$, $\Phifar$ & total azimuthal sweep on the near and far branches \\
$\Znear$, $\Zfar$ & total axial travel on the near and far branches \\
$\Phi$, $Z$ & sweep and axial travel of whichever branch is being solved \\
$\Wnear$, $\Wfar$ & branch functions whose negative gradients give $(\Phi,Z)$ \\
$J$ & Jacobian of $(\Phi,Z)$ with respect to $(\Lam,C_z)$ \\
$r_c$ & object radius at which the limiting medium changes, at fixed $C_z$ \\
$Z_0$, $C_{\max}$, $C_\pi$ & radial-ray axial travel, maximum feasible tilt, and onset tilt for sweeps beyond $\pi$ \\
$\theta_i$, $H_i$, $L_i$, $L_{\mathrm{tot}}$ & angle to the normal, normal thickness, lateral travel in medium $i$, and total lateral offset (flat window only) \\
\bottomrule
\end{tabular}
\end{table}

\subsection{Conserved quantities}
\label{sec:invariants}

Forward projection is equivalently solving which ray direction from $\va*{P}_0$ reaches $\va*{X}$ after refraction. The ray direction has two angular degrees of freedom, but the interface geometry may fix the plane of the ray and leave only one unknown. For the remaining freedom(s), we can use conserved quantities as parameters.

At an interface with unit normal $\nrm$, vector Snell's law states that the tangential part of $n\dvec$ is continuous:
\begin{equation}
\label{eq:snell_vec}
    n_a\qty\big(\dvec_a - (\dvec_a\vdot\nrm)\nrm) = n_b\qty\big(\dvec_b - (\dvec_b\vdot\nrm)\nrm) ,
\end{equation}
where $a$ and $b$ label the media on either side and $\dvec_a$, $\dvec_b$ are the corresponding unit ray directions. Equivalently,
\begin{equation}
    n_a \dvec_a - n_b \dvec_b = \qty\big[n_a(\dvec_a\vdot\nrm) - n_b(\dvec_b\vdot\nrm)]\nrm, \label{eq:nd_conserve}
\end{equation}
so refraction changes $n\dvec$ only in the normal direction. Along a straight segment in a homogeneous medium, both $n\dvec$ and $n\,\va*{x}\cross\dvec$ are constant.

For parallel flat interfaces, the normal has the same direction everywhere, so the tangential part of $n\dvec$ is conserved as a vector. It points along the known lateral displacement from $\va*{P}_0$ to $\va*{X}$. The only unknown is its magnitude $\Lam=n\sin\theta$, where $\theta$ is the angle to the normal direction. This is therefore the single unknown used in Section~\ref{sec:flat}.

For coaxial cylinders, the normal is radial, $\nrm=(\cos\varphi,\sin\varphi,0)$ with $\varphi$ as the azimuth angle. The axial direction is tangent to every interface, so Eq.~\eqref{eq:snell_vec} conserves $nd_z$. The azimuthal direction also lies in the tangent plane, but changes with $\varphi$. Thus, the quantity conserved along the whole ray is instead the $z$ component of $n\,\va*{x}\cross\dvec$, where $\va*{x}=(r\cos\varphi,r\sin\varphi,z)$ is any point on the ray (in cylindrical coordinates). The reason is that when crossing an interface, $n_a\dvec_a-n_b\dvec_b$ is radial, so its cross product with $\va*{x}$ has no $z$ component. In summary, the two conserved scalars are
\begin{equation}
    C_z\equiv nd_z, \quad \Lam \equiv n (xd_y - yd_x)=nr d_\varphi, \label{eq:invariants}
\end{equation}
named as the axial momentum and the angular momentum about the axis. The scalar $d_\varphi=\dvec\vdot\vu*{\varphi}$ is the component along the local azimuthal direction drawn in Fig.~\ref{fig:geometry}(a), where $\vu*{\varphi}=(-\sin\varphi,\cos\varphi,0)$.
Since a cylindrical ray can advance both around and along the axis, projection must solve for both $C_z$ and $\Lam$ to match the axial and azimuthal offsets of $\va*{X}$.

For concentric spheres, two unknowns in the cylinders' case are reduced to one because the rays stay in a fixed plane. At each spherical interface, the normal is parallel to $\va*{x}$. The entire vector $n\,\va*{x}\cross\dvec$ is conserved based on Eq.~\eqref{eq:nd_conserve}, which is normal to the plane that $O$, $\va*{P}_0$ and $\va*{X}$ lie in. Therefore, we only need to solve for the magnitude $\Lam=n\,r\sin\theta$, where $\theta$ is the angle between $\va*{x}$ and $\dvec$. This is also known as the Bouguer's invariant~\cite{born1999principles}. 

In summary, forward projection requires one scalar for a flat window, two for a cylinder, and one for a sphere. The conserved momenta themselves are classical results of Hamiltonian optics~\cite{born1999principles}. They also describe skew rays in circular optical fibres~\cite{snyder1983optical} and rays in a spherically layered Earth~\cite{aki2002quantitative}. Our contribution is the forward projection algorithm and visibility analysis built from them.

\section{Flat windows in invariant form}
\label{sec:flat}

To the best of our knowledge, there are two papers deriving the closed-form equations for the forward projection through refractive flat windows.
The first paper, from Agrawal et al.~\cite{agrawal2012theory}, showed that the forward projection through a flat window is a root of a polynomial, of degree 12 for a thick window between air and water. Then, Gu et al. replaced the polynomial by a single equation with one unknown, which can be solved by Newton's method in only 4--5 iterations~\cite{gu2023refraction}. Here we restate their equation in terms of the invariant $\Lam$ of Section~\ref{sec:invariants}.

Suppose there are $N$ media, as shown in Fig.~\ref{fig:index}(a), and $H_i$ is the thickness of medium $i$ ($1\leq i\leq N$) along the normal direction, the lateral travel in medium $i$ is $L_i=H_i\tan\theta_i$. Note that $H_1$ and $H_N$ are the distances from the pinhole $\va*{P}_0$ or the object point $\va*{X}$ to the nearest interface.
With $\Lam=n_i\sin\theta_i$ as the major variable, the forward projection solves
\begin{equation}
\label{eq:flat_f}
    f(\Lam) = \sum_{i=1}^N L_i = \sum_{i=1}^N \frac{H_i\Lam}{\sqrt{n_i^2-\Lam^2}} = L_{\text{tot}},
\end{equation}
where $L_{\text{tot}}$ is the lateral offset between the pinhole and the point. By replacing $k_i=1/n_i$ and $a=\Lam$, the above equation is the same as the equation showed in \cite{gu2023refraction}.
As Gu et al. noted, the root to this equation must exist and be unique. The reason is that: 
each term of $f$ increases with $\Lam$, due to $f'>0$; 
$f(0)=0$, and $f\to\infty$ as $\Lam\rightarrow\min_i n_i$, so every $L_{\text{tot}} \geq 0$ is reached by exactly one ray.

To extend the flat-window uniqueness argument to the two unknowns of a cylindrical window, we first express the projection equation as a derivative of one scalar, $W(\Lam)$, defined as
\begin{equation}
\label{eq:flat_W}
    W(\Lam) \equiv \sum_{i=1}^N H_i\sqrt{n_i^2-\Lam^2} ,
    \qquad
    f(\Lam) = -\dv{W}{\Lam} .
\end{equation}
Each term of $W$ is the layer thickness $H_i$ multiplied by the normal component $n_i\cos\theta_i$ of the ray momentum. Since $W''=-f'<0$, its strict concavity is the monotonicity that made the flat-window solution unique.

For a cylindrical window, each trial pair $(\Lam,C_z)$ predicts how far the ray turns around the axis and advances along it. Forward projection solves for the pairs that match the offsets from $\va*{P}_0$ to $\va*{X}$. Changing $\Lam$ or $C_z$ can also change the axial advance or the azimuthal sweep. It means that knowing the properties of $\pdv*{f}{\Lam}$ or $\pdv*{f}{C_z}$ along do not guarantee a unique solution. However, strict concavity of $W_{\text{near}}$ on the convex domain of momentum pairs that can reach the object's radius ensures that two different pairs cannot produce the same offsets. As shown in Section~\ref{sec:near}, a near-branch ray is therefore unique whenever it exists.
In the following, we first derive the cylindrical projection equations and show how both offsets follow from the corresponding scalar function in Section~\ref{sec:equations}, and then use the concavity of $W_{\text{near}}$ to establish near-branch uniqueness in Section~\ref{sec:near}. Finally, we check the properties of the far branch, where multiple rays can occur, in Section~\ref{sec:visibility}.

\section{Closed-form projection equations}
\label{sec:equations}
In this section, we derive the projection equations for coaxial cylinders. Section~\ref{sec:sphere} simplifies them to spheres. The derivation is based on the invariants of Eq.~\eqref{eq:invariants} and elementary geometry, and it ends up with two closed-form equations for $(\Lam,C_z)$.

\subsection{Propagation through one medium}
\label{sec:propagation}

\begin{figure}[htbp]
\centering
\begin{tabular}{@{}c@{}}
\paperfig[0.86\linewidth]{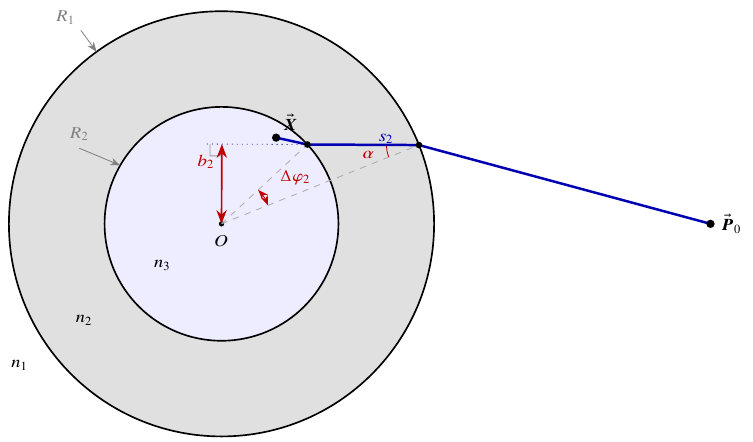} \\ (a) \\[5mm]
\paperfig[0.82\linewidth]{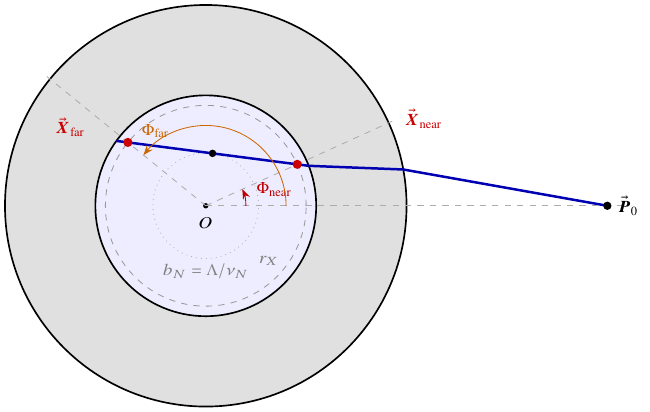} \\ (b)
\end{tabular}
\caption{Transverse geometry of a cylindrical ray (schematic). (a) In medium $2$, the chord from $R_1$ to $R_2$ has impact parameter $b_2=\Lam/\nu_2=R_1\sin\alpha$, where $\alpha$ is its angle to the inward radius at $R_1$. Its length and azimuthal sweep are $s_2$ and $\Delta\varphi_2$. (b) At fixed $(\Lam,C_z)$, the ray meets the object circle $r=r_X$ at $\va*{X}_{\mathrm{near}}$ on its inward leg and at $\va*{X}_{\mathrm{far}}$ after closest approach (dotted circle, radius $b_N=\Lam/\nu_N$). The corresponding sweeps from $\va*{P}_0$ are $\Phinear$ and $\Phifar$.}
\label{fig:transverse}
\end{figure}

In medium $k$, the axial component of $\dvec$ is $d_z=C_z/n_k$, so the transverse part of $\dvec$ has length $\norm{\dvec_\perp}=\nu_k/n_k$, with
\begin{equation}
\label{eq:nu}
    \nu_k \equiv \sqrt{n_k^2-C_z^2} ,
\end{equation}
which requires $\abs{C_z}<\min_kn_k$. If we project the ray onto a plane $z=\text{const}$, the ray in medium $k$ is still a straight line, and we can define the distance between the central axis and the ray as $b_k$. For example, in Fig.~\ref{fig:transverse}(a), $b_2$ is the distance between the ray in medium 2 and the axis point $O$. $b_k$, named as the impact parameter in medium $k$, can be derived as
\begin{equation}
\label{eq:impact}
    b_k = r\sin\alpha = r\,\frac{d_\varphi}{\norm{\dvec_\perp}} = r\,\frac{\Lam/(n_kr)}{\nu_k/n_k} = \frac{\Lam}{\nu_k} ,
\end{equation}
where $r$ is the radial distance of a point on the ray within medium $k$, and $\alpha$ is the angle between the transverse direction $\dvec_\perp$ and the inward radial direction. 

As shown in Fig.~\ref{fig:transverse}(a), the ray travelling in medium 2 changes its azimuth angle as $\Delta\varphi_2 = \arcsin(b_2/R_2) - \arcsin(b_2/R_1)$. For a general azimuth angle change in medium $k$, travelling from $R_{k-1}$ to $R_k$, we have
\begin{equation}
\label{eq:sweep}
    \Delta\varphi_k = \arcsin\frac{b_k}{R_k} - \arcsin\frac{b_k}{R_{k-1}} \geq 0
\end{equation}
, and the axial change is calculated as 
\begin{gather}
s_k = \sqrt{R_{k-1}^2 - b_k^2} - \sqrt{R_k^2 - b_k^2} \label{eq:chord} , \\
\Delta z_k = \frac{d_z}{\norm{\dvec_\perp}} s_k = \frac{C_z}{\nu_k}s_k \label{eq:dz},
\end{gather}
where $s_k$ is the transverse chord length in medium $k$.

\subsection{Admissible domain for $\Lam$}
\label{sec:admissible}

In this part, we want to discuss the requirements for a ray that can cross the interface between media $k$ and $k+1$.  
The first requirement is that its transverse radial distance must be smaller than $R_k$, otherwise the ray cannot reach the interface. It means $b_k<R_k$, i.e. $\Lam<\nu_kR_k$. By applying this to every medium, we can get the following requirement
\begin{equation}
\label{eq:adm}
    0 \le \Lam < \Lmax \equiv \min_{1\le k\le N}\nu_kR_k \equiv \nu_{k^*}R_{k^*},
\end{equation}
where $R_N\equiv r_X$ and $k^\ast$ is the medium that reaches the minimum. 

The second requirement is that the ray must not be total reflected. It means that $n_k\sin\theta_k<n_{k+1}$, where $\theta_k$ is the incident angle onto the tangent plane of the cylindrical interface. Then, we can get 
\begin{gather}
    \sin^2\theta_k = d_\varphi^2 + d_z^2 = \frac{\Lam^2}{n_k^2 R_k^2} + \frac{C_z^2}{n_k^2}, \\ 
    n_k^2\sin^2\theta_k = \frac{\Lam^2}{R_k^2} + C_z^2 < n_{k+1}^2 \Leftrightarrow \Lam<\nu_{k+1}R_k.
\end{gather}
Note that Eq.~\eqref{eq:adm} already satisfies the total reflection criterion, since $\Lam<\nu_{k+1}R_{k+1}<\nu_{k+1}R_k$. 

In summary, reaching the next radius sets the admissible domain of $\Lam$, and the total reflection imposes no additional restriction. This differs from the flat window, where total reflection is the main constraint (Section~\ref{sec:flat}).

\subsection{Two branches}
\label{sec:branches}

When the ray travelling through the media, its transverse radial distance monotonically decreases ($d_r<0$), except in the last medium $N$ where $\va*{X}$ lies in. 
In the last medium $N$, it may pass its closest point to the axis, at $r=b_N$, and meet the circle $r=r_X$ a second time on its way out, as shown in Fig.~\ref{fig:transverse}(b). 
We name the two cases as: (1) the near branch, where the ray reaches $r=r_X$ before its closest point; (2) the far branch, where the ray reaches $r=r_X$ afterwards. 

By summing Eq.~\eqref{eq:sweep} over the media, we can get the expression of the total amzimuth sweep, from $\va*{P}_0$ to $\va*{X}$, for each branch as
\begin{align}
\label{eq:Phinear}
    \Phinear(\Lam,C_z) &= \sum_{k=1}^{N}\qty[\arcsin\frac{\Lam}{\nu_kR_k} - \arcsin\frac{\Lam}{\nu_kR_{k-1}}] , \\
\label{eq:Phifar}
    \Phifar(\Lam,C_z) &= \Phinear(\Lam,C_z) + 2\arccos\frac{\Lam}{\nu_Nr_X} ,
\end{align}
with $\Phinear(0, C_z)=0$ and $\Phifar(0, C_z)=\pi$. 
Summing Eq.~\eqref{eq:dz} gives the axial offset on the near branch. On the far branch, the ray travels from $r_X$ to its closest approach and back, adding an axial offset in medium $N$:
\begin{align}
\label{eq:Z}
    \Znear(\Lam,C_z) &= C_z\sum_{k=1}^{N}\frac{\sqrt{\nu_k^2R_{k-1}^2-\Lam^2}-\sqrt{\nu_k^2R_k^2-\Lam^2}}{\nu_k^2} , \\
\label{eq:Zfar}
    \Zfar(\Lam,C_z) &= \Znear(\Lam,C_z)+\frac{2C_z}{\nu_N^2}\sqrt{\nu_N^2r_X^2-\Lam^2} .
\end{align}

In the projection equations below, $(\Phi,Z)$ denotes $(\Phinear,Z_{\text{near}})$ or $(\Phifar,Z_{\text{far}})$ for the branch being solved. Projecting $\va*{X}$ means solving
\begin{equation}
\label{eq:system}
    \Phi(\Lam,C_z) = \Delta\varphi ,
    \qquad
    Z(\Lam,C_z) = \Delta z \equiv z_X - z_P
\end{equation}
on each branch, where $\Delta\varphi\in[0,\pi]$ is the azimuth angle between the projections of $\va*{P}_0$ and $\va*{X}$ on the transverse plane. 
The restriction to $\Delta\varphi\le\pi$ assumes that no ray sweeps more than $\pi$ before it reaches $\va*{X}$, but that could still happen and the method to handle that scenario will be discussed in Section~\ref{sec:wrap}.

\subsection{Scalar function representation}
\label{sec:action}

On each branch, the azimuthal sweep and axial offset can be represented as derivatives of one scalar function. For the near branch, we can define
\begin{equation}
\label{eq:W}
    W_{\text{near}}(\Lam,C_z) \equiv \sum_{k=1}^{N}\int_{R_k}^{R_{k-1}}\sqrt{\nu_k^2-\frac{\Lam^2}{r^2}}\,\dd r ,
\end{equation}
whose integrand is the radial component of the ray momentum. It is the analogue of Eq.~\eqref{eq:flat_W}, by replacing each layer thickness $H_i$ with an integral over the radial extent of medium $k$.
For the far branch, the ray passes the turning point $b_N=\Lam/\nu_N$ and returns to $r_X$. Since the extra radial interval is traversed twice, it gives
\begin{equation}
\label{eq:Wfar}
\begin{aligned}
    W_{\text{far}}(\Lam,C_z)
    &= W_{\text{near}}(\Lam,C_z)+2\int_{b_N}^{r_X}\sqrt{\nu_N^2-\frac{\Lam^2}{r^2}}\,\dd r \\
    &= W_{\text{near}}(\Lam,C_z)+2\qty[\sqrt{\nu_N^2r_X^2-\Lam^2}-\Lam\arccos\frac{\Lam}{\nu_Nr_X}] .
\end{aligned}
\end{equation}
The negative derivatives of these functions then give the ray's azimuthal sweep and axial offset on each branch:
\begin{align}
\label{eq:gradW}
    -\grad W_{\text{near}} &= (\Phinear,Z_{\text{near}}) , \\
\label{eq:gradWfar}
    -\grad W_{\text{far}} &= (\Phifar,Z_{\text{far}}) .
\end{align}

Because both ray offsets come from the same branch-specific function, their responses to the two momenta $(\Lam,C_z)$ are linked. The sensitivity of the azimuthal sweep to $C_z$ equals the sensitivity of the axial offset to $\Lam$, so the Jacobian is symmetric:
\begin{equation}
\label{eq:symmetry}
    \pdv{\Phi}{C_z} = \pdv{Z}{\Lam}
    = C_z\Lam\sum_{k=1}^{N}\frac{\q{\nu_kR_k}-\q{\nu_kR_{k-1}}}{\nu_k^2}
    \;-\; \underbrace{\frac{2C_z\Lam\,\q{\nu_Nr_X}}{\nu_N^2}}_{\text{far branch only}} ,
\end{equation}
where $q(a)\equiv1/\sqrt{a^2-\Lam^2}$.
The above equation will also be used in Section~\ref{sec:tilt} to determine how the axial offset changes with tilt at a fixed azimuth. 

\subsection{Map from $(\Lam,C_z)$ to pixel}
\label{sec:pixel}

Once $(\Lam,C_z)$ is found, Eq.~\eqref{eq:invariants} gives two components of the ray direction at the pinhole:
\begin{equation}
\label{eq:pinhole_direction}
    d_z=\frac{C_z}{n_1},\qquad d_\varphi=\frac{\Lam}{n_1r_P},\qquad
    d_r=-\sqrt{1-d_z^2-d_\varphi^2} ,
\end{equation}
where $d_r$ is chosen with a negative sign, because we want to select a ray that initially travels inward from $\va*{P}_0$.
The full direction of the ray through $\va*{P}_0$ is therefore $\dvec=d_r\vu*{r}+d_\varphi\vu*{\varphi}+d_z\vu*{z}$.
The regular pinhole model then maps the direction to a pixel, where the camera intrinsic and extrinsic parameters are needed.

\section{The near-branch solver}
\label{sec:near}

In Section~\ref{sec:equations}, we reduce the cylindrical projection problem to solving for $(\Lam,C_z)$.
In this section, we focus on the near-branch projection. First, we will prove that the solution is unique whenever it exists, and then describe how to solve for it within the admissible domain.

\subsection{Uniqueness of near-branch projection}
\label{sec:convex}

\begin{figure}[htbp]
\centering
\paperfig[0.70\linewidth]{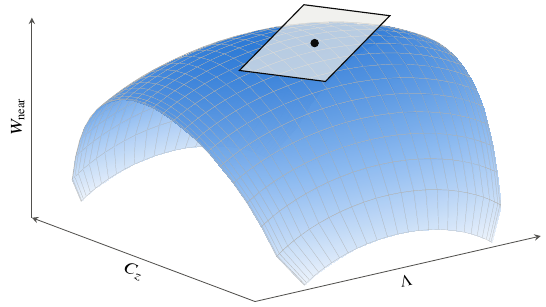}
\caption{Near-branch action $\Wnear(\Lam,C_z)$ for the reference cylindrical window at $r_X=50$\,mm. The pale plane is tangent at the marked momentum pair. Its slopes are $-\Phinear$ along $\Lam$ and $-\Znear$ along $C_z$; for a ray reaching the target, these equal $-\Delta\varphi$ and $-\Delta z$. Strict concavity permits at most one tangent point with the target's slopes. $\Lam$ and $\Wnear$ are in millimetres; $C_z$ is dimensionless.}
\label{fig:cyl_dome}
\end{figure}

Each point $(\Lam,C_z)$ on the surface in Fig.~\ref{fig:cyl_dome} represents a ray launched from the pinhole: $\Lam$ sets its azimuthal component and $C_z$ its axial component. By Eq.~\eqref{eq:gradW}, the negatives of the surface slopes along $\Lam$ and $C_z$ equal the ray's total azimuthal sweep and axial offset. The target offsets $(\Delta\varphi,\Delta z)$ therefore specify the slopes of a tangent plane. Because the near-branch surface bends downward in every direction, at most one momentum pair can have those slopes. Two properties, as shown below, make this argument precise: the momentum pairs of near-branch rays form a convex domain, and the surface bends downward everywhere in that domain.

First, we show that the admissible domain is convex.
From Section~\ref{sec:admissible}, a ray can reach the radius $R_k$ only if $\abs{\Lam}<\nu_kR_k$.
Since $\nu_k^2=n_k^2-C_z^2$, the admissible domain for momentum pairs of near-branch rays can be defined as
\begin{equation}
\label{eq:domain}
    \mathcal{D}=\{(\Lam,C_z): C_z^2+\Lam^2/R_k^2<n_k^2 \text{ for } 1\le k\le N\},
\end{equation}
with $R_N=r_X$.
Each of the $N$ conditions describes the inside of an ellipse in the $(\Lam,C_z)$ plane.
Therefore, $\mathcal{D}$ is convex:
a straight segment between two points inside an ellipse stays inside it, so a segment between two points of $\mathcal{D}$ stays inside all $N$ ellipses.

Second, we show that $\Wnear$ bends downward everywhere (concavity) in $\mathcal{D}$.
At a fixed radius $r$ in medium $k$, the integrand of Eq.~\eqref{eq:W} is the radial component of the ray momentum, $h=\sqrt{n_k^2-\vb*{v}^{\top}Q\,\vb*{v}}$, where $\vb*{v}=(\Lam,C_z)^\top$ and $Q=\operatorname{diag}(1/r^2,1)$.
Its Hessian with respect to $\vb*{v}$, $\grad^2h=-Q/h-Q\vb*{v}\vb*{v}^{\top}Q/h^3$, is negative definite, because $Q$ is positive definite and $h>0$, since $r\ge R_k$ throughout medium $k$.
Integrating over the radial extent of each medium and summing over the media keeps the Hessian negative definite, so $\Wnear$ is strictly concave everywhere in $\mathcal{D}$, as shown in Fig.~\ref{fig:cyl_dome}.

Together, these two properties give the uniqueness: along the straight segment joining two different pairs in $\mathcal{D}$, the slope of $\Wnear$ strictly decreases, so the two pairs cannot have the same slopes, i.e. cannot reach the same target as defined by Eq.~\eqref{eq:gradW}.
These properties also make the Jacobian of $(\Phinear,\Znear)$, $J=-\grad^2\Wnear$, positive definite on $\mathcal{D}$. 
Thus, $\Phinear$ increases with $\Lam$ and $\det J>0$, which is used in Sections~\ref{sec:newton} and~\ref{sec:visibility}.

\subsection{Newton's method}
\label{sec:newton}

Section~\ref{sec:convex} shows that a near-branch solution is unique if it exists. Uniqueness alone does not establish visibility: a Newton iteration started from an arbitrary pair may fail because the target is invisible or because the starting pair is poor. The reach and axial-offset tests in Sections~\ref{sec:tilt} and~\ref{sec:algorithm} determine whether a ray exists and provide a starting pair $(\Lam,C_z)$. Newton's method then refines both momenta together. Let $\vb*{v}=(\Lam,C_z)^\top$ and define the two projection errors and their Jacobian as
\begin{equation}
\label{eq:newton_residual}
    \vb*{F}(\vb*{v})=
    \begin{pmatrix}
        \Phinear-\Delta\varphi \\
        Z_{\text{near}}-\Delta z
    \end{pmatrix},
    \qquad
    J(\vb*{v})=\pdv{\vb*{F}}{\vb*{v}} .
\end{equation}
At each iteration, e.g. at iteration $n$, solve for the correction $\vb*{\delta}$ and update both momenta:
\begin{equation}
\label{eq:newton_step}
    J(\vb*{v}^n)\,\vb*{\delta}^n=-\vb*{F}(\vb*{v}^n),
    \qquad
    \vb*{v}^{n+1}=\vb*{v}^n+t^n\vb*{\delta}^n .
\end{equation}
In order to keep $\vb*{v}$ stay in the admissible domain $\mathcal{D}$ with $\Lam \geq 0$, we first try $t^n=1$ and shorten the step if needed. The Jacobian is positive definite on $\mathcal{D}$ (Section~\ref{sec:convex}), so the Newton correction is well defined there.

Convergence is determined based on the projection errors. To compare the axial error, which has units of length, with the angular error, we divide it by a scale close to $r_P$. The implementation uses $l=2^{\lfloor\log_2 r_P\rfloor}$ rather than $r_P$ itself. Because division by the power of two introduces no rounding in binary floating point, which matters near grazing incidence where the ray sweep is especially sensitive to small changes in momentum. 
The residual is then defined as
\[
\epsilon=\sqrt{(\Phinear-\Delta\varphi)^2+\qty[\frac{Z_{\text{near}}-\Delta z}{l}]^2} ,
\]
which will trigger stopping when it falls below the chosen tolerance.

\section{Visibility and multiple images}
\label{sec:visibility}

Section~\ref{sec:near} shows that the near-branch equations have at most one solution for a prescribed angular sweep and axial offset. But it does not determine whether that solution exists or whether the far branch provides additional images.
In this section, we first find the near branch's limiting angular sweep and the radius at which the two branches separate, for a fixed tilt $C_z$.
Then we compare the angular ranges of the near and far branches to see which azimuths none of them reach, and which both reach. The folds of the far-branch sweep, where two far-branch rays can reach the same azimuth offset, are also located.
Finally, we impose the target's axial offset and account for rays sweeping beyond $\pi$ to determine visibility and image count in three dimensions.

\subsection{Angular reach and far-branch folds}
\label{sec:angularreach}

In this subsection, we focus on how to determine the reachable target azimuths and how many images (solutions) each target azimuth would have.
In order to reach the target azimuth, a ray must first pass every cylindrical interface. At fixed tilt $C_z$, this requirement limits its angular momentum $\Lam$.
Thus, we use that limit to compare the near- and far-branch sweeps, and then account for the target's axial offset in Section~\ref{sec:tilt}.

As shown in Section~\ref{sec:admissible}, a ray reaches radius $R_k$ only if $b_k<R_k$, i.e. $\Lam<\nu_kR_k$ ($1\leq k\leq N$). At fixed $C_z$, the interface radii are fixed while the object radius $R_N\equiv r_X$ varies with the target. The limiting value of $\Lam$ switches from the object circle to an earlier interface at the critical radius
\begin{equation}
    \label{eq:rc}
    r_c \equiv \frac{\min_{k<N}\nu_k R_k}{\nu_N} .
\end{equation}
When $r_X \leq r_c$, the object circle alone limits $\Lam$ with $\Lmax=\nu_Nr_X$. The near and far sweeps will meet as $\Lam\to\Lmax$ because the extra far-branch sweep in Eq.~\eqref{eq:Phifar} vanishes.
When $r_X>r_c$, an earlier interface limits $\Lam$, so the two sweeps remain separate at $\Lmax$.

For the reference window of Section~\ref{sec:setup}, Eq.~\eqref{eq:rc} gives $r_c=75.19$\,mm for rays in the transverse plane ($C_z=0$).

\begin{figure}[htbp]
\centering
\paperfig[\linewidth]{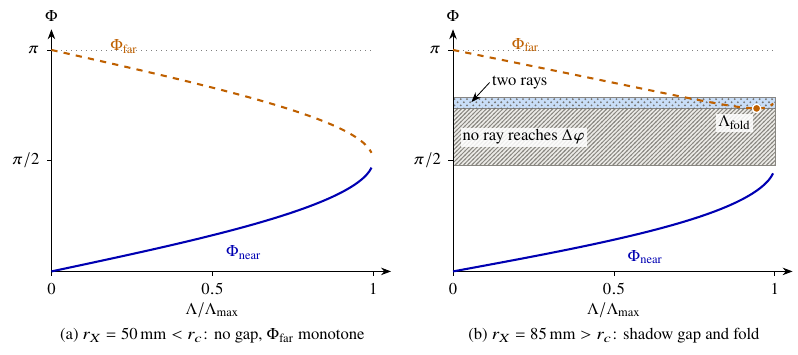}
\caption{Azimuthal sweep $\Phi$ versus $\Lam/\Lmax$ for the reference window at $C_z=0$ ($r_c=75.19$\,mm). Solid and dashed curves denote near and far branches; intersections with a target level $\Delta\varphi$ give rays to a point in the pinhole plane. (a) At $r_X=50$\,mm$<r_c$, the object circle limits $\Lam$; the branches meet at grazing and together cover $[0,\pi]$ once. (b) At $r_X=85$\,mm$>r_c$, the outer interface limits $\Lam$. The hatched band between the near-branch limit and far-branch minimum is unreachable. The dotted band, between that minimum at $\Lam_{\text{fold}}$ and the far-branch endpoint, has two far-branch rays.}
\label{fig:phi}
\end{figure}

Fig.~\ref{fig:phi} shows the comparison between objects at radius $r_X=50$\,mm (inside $r_c$) and $85$\,mm (outside $r_c$). In panel (a), the curves of near and far branches meet at the grazing limit. In panel (b), there is an angular range without visible ray, because the outer interface limits $\Lam$ and $\Lam$ cannot reach $\nu_N r_X$. Besides the angular gap, we can also find that the dashed far-branch curve bends back after $\Lam_{\text{fold}}$, which means a horizontal line can cross it twice, i.e. multiple projected images for $\va*{X}$ could coexist.

To quantify the gap seen in Fig.~\ref{fig:phi}(b), we compare the full angular ranges of the two branches. As shown in Section~\ref{sec:convex}, the near-branch sweep increases with $\Lam$, so its largest value is $\Phinear(\Lmax)$.
When $r_X>r_c$, the far-branch sweep can instead decrease and then increase, reaching its minimum before $\Lmax$.
If that minimum exceeds $\Phinear(\Lmax)$, no ray with $\Phi\leq\pi$ reaches a target azimuth in the interval
\begin{equation}
\label{eq:gap}
    \Phinear(\Lmax) \;<\; \Delta\varphi \;<\; \Phi_{\min},
    \qquad \Phi_{\min} \equiv \min_{0\le\Lam\le\Lmax}\Phifar(\Lam) \;\le\; \pi .
\end{equation}
If $\Phi_{\min}\le\Phinear(\Lmax)$, the branch ranges meet or overlap, and the angular gap will not appear (Supplement~1, Section~S6).

To locate the far-branch minimum in Eq.~\eqref{eq:gap}, we examine how each sweep changes with $\Lam$. Differentiation at fixed $C_z$ gives
\begin{equation}
\label{eq:dPhi}
    \Phinear' = S + \q{\nu_Nr_X} - \q{\nu_1r_P} ,
    \qquad
    \Phifar' = S - \q{\nu_Nr_X} - \q{\nu_1r_P} ,
\end{equation}
where
\begin{equation}
\label{eq:S}
    S \equiv \sum_{k=1}^{N-1}\qty\big[\q{\nu_kR_k}-\q{\nu_{k+1}R_k}] .
\end{equation}
The sum $S$ collects the changes in slope at the interfaces. Each bracket is positive when the ray enters a denser medium and vanishes when the refractive indices match. The other terms come from the object and pinhole. If $\Phifar'$ changes from negative to positive, the far-branch sweep reaches a minimum. Rays on either side of that minimum can then reach the same azimuth. As the target radius varies, the minima trace a caustic.

\begin{figure}[htbp]
\centering
\paperfig[\linewidth]{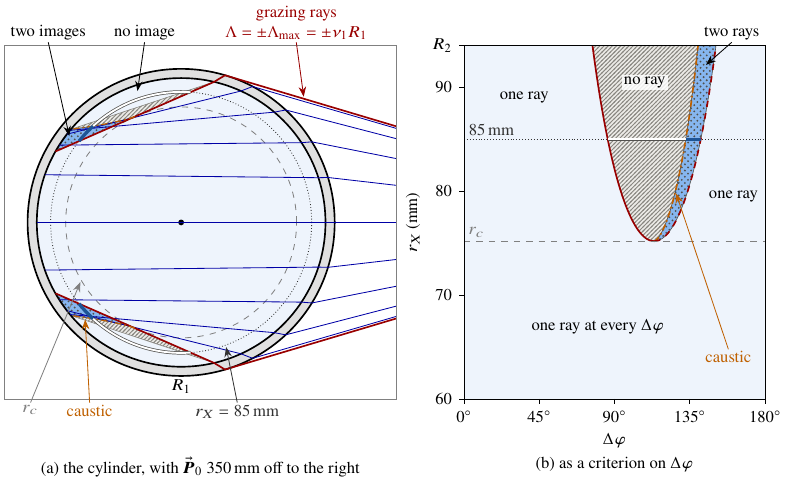}
\caption{Visibility in the pinhole plane of the reference window ($C_z=0$). (a) Transverse view: thin lines are sample rays from $\va*{P}_0$, located to the right of the panel; thick lines graze the outer wall at $\Lam=\pm\Lmax=\pm\nu_1R_1$. The dash-dotted caustic separates the hatched no-image lobes from adjacent dotted two-image bands; the dashed circle marks $r_c$. (b) The same regions in $(\Delta\varphi,r_X)$: the near-branch grazing limit $\Phinear(\Lmax)$ (solid) and caustic $\min\Phifar$ (dash-dotted) bound the no-image band; the far-branch grazing limit $\Phifar(\Lmax)$ (dashed) closes the two-image band. At $r_X=85$\,mm, their azimuthal ranges are $85.9^\circ$--$132.5^\circ$ and $132.5^\circ$--$141.5^\circ$, respectively.}
\label{fig:shadow}
\end{figure}

For the example case introduced above, the far-branch sweep decreases throughout its domain when $r_X\le r_c$, so the two branches cover $[0,\pi]$ once. At $r_X=85$\,mm, the far branch instead has the minimum shown in Fig.~\ref{fig:phi}(b).
Fig.~\ref{fig:shadow} further illustrates the meaning of these sweep limits. In panel (a), near-branch rays grazing the outer wall bound one side of the invisible regions. The minima of the far-branch sweep trace the caustic along their other side. Between the caustic and the outbound grazing rays, two far-branch rays reach the same point in the dotted regions. 
Panel (b) shows the same boundaries in the $(\Delta\varphi,r_X)$ plane. 

Such shadow and multiple-image zones are also observed by Paolillo and Astarita numerically in a PMMA cylinder, with similar air--acrylic--water interfaces~\cite{paolillo2022perspective}. But they did not provide a complete analysis on when and why those regions appear. In the following, we will provide a complete algorithm that can determine different regions.

The behavior in Fig.~\ref{fig:shadow} is not determined by $r_c$ alone. Although $r_X>r_c$ means that an interface limits $\Lam$, it does not guarantee a fold in $\Phifar$ or an angular gap: the far branch may remain monotone, and the angular ranges of the two branches may overlap (Supplement~1, Sections~S2 and S6). Conversely, a multilayer cylinder can have far-branch folds even when $r_X<r_c$. In such case, two folds might give three far-branch rays to the same point in the pinhole plane (Supplement~1, Section~S6). 
We therefore examine the full far-branch sweep to count the rays reaching each azimuth. Supplement~1, Section~S2 gives sufficient conditions for a monotone far branch and bounds for locating all its folds.

\subsection{Visibility across ray tilts}
\label{sec:tilt}

The above analysis shows the reachable azimuths for a ray at fixed tilt $C_z$. A 3-D target also has an axial offset $\Delta z$, so a ray even with the correct azimuth may still pass above or below it. Therefore, we must vary $C_z$ and match both offsets. This changes $\nu_k=\sqrt{n_k^2-C_z^2}$ and hence the allowed range of $\Lam$ and the angular reach.

For each tilt that reaches the target azimuth, let $\Lam(C_z)$ denote the solution to $\Phi(\Lam(C_z),C_z)=\Delta\varphi$.
When $\pdv*{\Phi}{\Lam}\neq0$, we can calculate the total derivative of $Z$ on $C_z$ as
\begin{gather}
    \dv{C_z} \Phi(\Lam(C_z), C_z) = \pdv{\Phi}{\Lam}\pdv{\Lam}{C_z} + \pdv{\Phi}{C_z} = 0 \Rightarrow \pdv{\Lam}{C_z} = -\frac{\pdv*{\Phi}{C_z}}{\pdv*{\Phi}{\Lam}} ,\\
\label{eq:zetaprime}
    \dv{C_z}Z(\Lam(C_z), C_z) = \pdv{Z}{\Lam}\pdv{\Lam}{C_z} + \pdv{Z}{C_z} = \frac{\det J}{\pdv*{\Phi}{\Lam}} ,
\end{gather}
where $J$ is the Jacobian of $(\Phi,Z)$ with respect to $(\Lam,C_z)$. 
Eq.~\eqref{eq:zetaprime} basically means how will the change of $C_z$ affect the axial reach with a given azimuth reach. We will use this total derivative to guide the searching for $C_z$ to satisfy both requirements set by Eq.~\ref{eq:system}.

On the near branch, we know from Section~\ref{sec:convex} that $\pdv*{\Phinear}{\Lam}>0$ and $\det J>0$. Thus, as $C_z$ increases, the axial travel of the ray (already satisfies the azimuth criterion) also increases.
Once a continuous range of tilts that reaches $\Delta\varphi$ is known, we check whether the corresponding axial offsets include $\Delta z$. If so, $C_z$ will be found by 1D bisection.
To find all such tilts, we test whether the near branch's grazing-limit sweep satisfies $\Phinear(\Lmax(C_z);C_z)\ge\Delta\varphi$.
As $C_z$ varies, the medium that limits $\Lmax$ may also change. A change is possible only when $\nu_jR_j = \nu_kR_k$ ($j$ and $k$ are indices of distinct media, $1\le j<k\le N$). Solving this equality gives the candidate switch tilts
\begin{equation}
\label{eq:cuts}
    C_z^2 = \frac{n_j^2R_j^2-n_k^2R_k^2}{R_j^2-R_k^2} .
\end{equation}
The candidate tilts within the physical range divide the search into intervals where the same medium limits $\Lmax$. Within each interval, we bound the angular reach and subdivide as needed to identify the tilts that can reach $\Delta\varphi$ (Supplement~1, Section~S3).

The bisection search only works for near branch, because it relies on one value of $\Lam$ for each reachable tilt and an axial offset that increases with $C_z$. 
For far branch, a fold may exist, which can give multiple values of $\Lam$ at the same tilt and target azimuth. Thus, following only one $\Lam(C_z)$ curve could therefore miss some rays. 
To find every far-branch image, the algorithm in Section~\ref{sec:algorithm} searches the two-dimensional $(\Lam,C_z)$ domain for pairs that match both target offsets.

\subsection{Long-way-round rays and image count}
\label{sec:wrap}

\begin{figure}[htbp]
\centering
\includegraphics[width=0.68\linewidth,trim={185pt 18pt 0pt 0pt},clip]{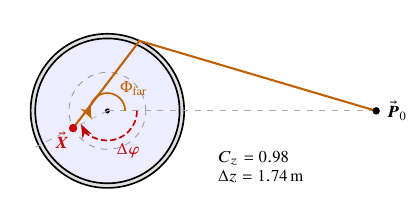}
\caption{Long-way-round far-branch ray through the reference window, shown in transverse projection. The solid ray from $\va*{P}_0$ to $\va*{X}$ sweeps $\Phifar=206.6^\circ=360^\circ-\Delta\varphi$; the dashed arc marks the shorter target separation $\Delta\varphi=153.4^\circ$ in the opposite azimuthal direction, not a second ray. Here $C_z=0.98$ and $\Delta z=1.74$\,m. Supplement~1, Section~S5 shows the full illustration and another example.}
\label{fig:wrapmain}
\end{figure}

The analysis above tests whether a ray can match the target's azimuthal separation $\Delta\varphi\in[0,\pi]$ and axial offset $\Delta z$.
However, a ray may also reach the same point after traveling around the other side of the cylinder. In the transverse view of Fig.~\ref{fig:wrapmain}, the far-branch ray sweeps $2\pi-\Delta\varphi$, while $\Delta\varphi$ marks the shorter angle to the target in the opposite sense of rotation. Other windows can even admit a ray that makes a further full turn, sweeping $\Delta\varphi+2\pi$ (Supplement~1, Section~S5). We therefore need to decide when to search these additional angular targets.

The axial offset provides a simple test for long-way-round rays. At fixed tilt, $\nu_k$ is fixed, and the inward chord in medium $k$ grows with its impact parameter $0\le b_k<R_k$. Differentiating Eq.~\eqref{eq:chord} gives
\begin{equation}
    \dv{s_k}{b_k}
    = b_k\qty(\frac{1}{\sqrt{R_k^2-b_k^2}}-\frac{1}{\sqrt{R_{k-1}^2-b_k^2}})\ge 0 \Rightarrow s_k\ge s_k(0)=R_{k-1}-R_k.
\end{equation}
On the far branch, the final chord passes the closest approach before reaching $\va*{X}$, hence the above inequality is still satisfied:
\begin{equation}
    s_N^{\mathrm{far}}
    =\sqrt{R_{N-1}^2-b_N^2}+\sqrt{r_X^2-b_N^2}
    \ge \sqrt{R_{N-1}^2-b_N^2} - \sqrt{r_X^2-b_N^2} \equiv s_N \ge R_{N-1}-r_X.
\end{equation}
According to Eq.~\ref{eq:dz}, we can get the range of axial offset magnitude as
\begin{equation}
\label{eq:z0}
    \abs{Z(\Lam,C_z)} \;\ge\; Z_0(\abs{C_z}) ,
    \qquad
    Z_0(C) \equiv C\sum_{k=1}^{N}\frac{R_{k-1}-R_k}{\nu_k(C)} .
\end{equation}
Here $Z_0(C)$ is the axial travel of the inward radial ray at tilt magnitude $C$, and increases with $C$.

To decide whether additional angular targets are needed, we compare the largest sweep of each branch with $\pi$ at the object's radius $r_X$. At each tilt, the near branch's largest sweep is its limit as $\Lam\to\Lmax$, while the far branch's largest sweep may occur at an endpoint or an interior turning point. If neither branch can exceed $\pi$ at any tilt, only $\Delta\varphi$ needs to be searched.

Otherwise, suppose $C_\pi$ is the onset tilt for sweeps beyond $\pi$. The sweep analysis in Supplement~1, Section~S5 rules out long-way-round rays at every $\abs{C_z}<C_\pi$, so any such ray must have $\abs{C_z}\ge C_\pi$. Eq.~\eqref{eq:z0} then gives $\abs{Z}\ge Z_0(C_\pi)$, equivalently, the condition
\begin{equation}
\label{eq:wraptest}
    \abs{\Delta z} \;<\; Z_0(C_\pi),
\end{equation}
rules out a long-way-round image.

\begin{figure}[htbp]
\centering
\paperfig[\linewidth]{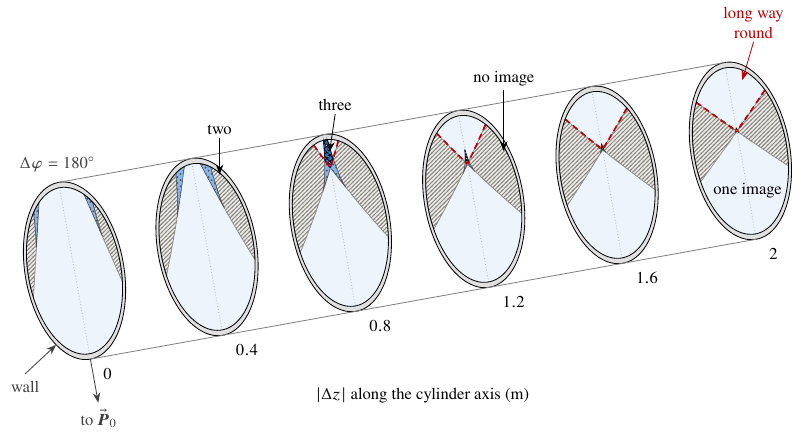}
\caption{Image count inside the reference window in transverse sections at $\abs{\Delta z}=0$, $0.4$, \dots, $2.0$\,m from the pinhole plane. Hatching marks no-image regions; the labeled bands have one, two, or three images. Red dashed boundaries mark the onset of long-way-round images. Each section is symmetric about its dotted diameter, the trace of the pinhole--axis plane. Section shapes are to scale, but axial spacing is compressed: $0.4$\,m is drawn as $1.5R_1$. Counts include both branches and wrapped angular targets, evaluated on a $0.25^\circ\times0.5$\,mm azimuth--radius grid.}
\label{fig:imgvol}
\end{figure}

After all relevant angles have been tested, each distinct ray matching both offsets gives one image. The near branch contributes at most one ray per angular target.
But the far branch may contribute multiple images due to the fold of $\Phifar$, or long-way-round rays when $\Phifar>\pi$. Different cases of the multiple images are discussed in details in Supplement~1, Section~S6.
To summarize different scenario mentioned above, the resulting image count maps are plotted in Fig.~\ref{fig:imgvol}. At the same transverse plane as $\va*{P}_0$, both invisible and two-image regions exist.
When going farther along the cylinder, three-image region also appears and one side of one-image region is due to the long way round rays.

These regions show that a point may produce no image or several images. Therefore, the algorithm shown in Section~\ref{sec:algorithm} will search for every ray reaching the point and returns the corresponding pixels, or an empty list when no ray exists.

\section{Projection algorithm}
\label{sec:algorithm}

In Section~\ref{sec:visibility}, we show that projecting a point requires finding every ray that matches its angular and axial offsets. The complete solution should include rays on both branches and rays travelling the long way round if exist. In this section, we are going to introduce our closed-form forward projection algorithm, which will provide complete solutions with efficiency.

\begin{figure}[htbp]
\centering
\paperfig[0.92\linewidth]{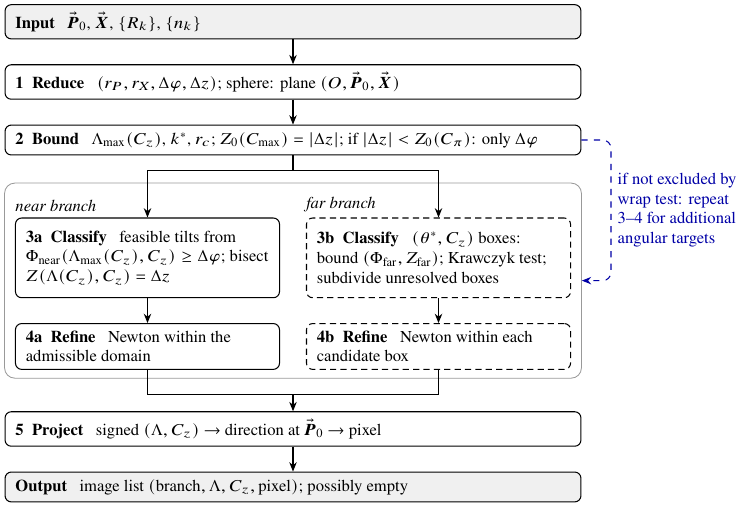}
\caption{Forward projection. The solid near-branch lane brackets feasible tilts and bisects the axial offset; the dashed far-branch lane bounds $(\Phi,Z)$ in boxes. Each candidate is refined by Newton's method, then its signed momenta determine the pinhole direction and pixel. If the wrap test~\eqref{eq:wraptest} does not exclude longer sweeps, both lanes are repeated for each additional angular target, replacing $\Delta\varphi$ in the near lane.}
\label{fig:pipeline}
\end{figure}

Our algorithm first bounds the possible $(\Lam,C_z)$ pairs, and then use the near branch's monotonicity or a bounded search on the far branch to locate solutions. Finally, each solution is refined and mapped to a pixel, as outlined in Fig.~\ref{fig:pipeline}.

In step (1), we express $\va*{X}$ by its radius $r_X$, azimuthal separation $\Delta\varphi$ and axial offset $\Delta z$ from the pinhole. Two cases are handled before the general search. If $r_X=0$, the target lies on the cylinder axis, so its azimuth is undefined and any ray reaching it has $\Lam=0$.
If $\Delta\varphi=\pi$, one possible far-branch ray passes through the cylinder axis in the transverse plane, so $\Lam=0$ and $\Phifar(0,C_z)=\pi$ at any tilt. Only its axial travel remains to be matched: we solve the monotone equation $\Zfar(0,C_z)=\Delta z$ for $C_z$ (Supplement~1, Section~S4). We add this ray directly to the candidate list because it lies on the $\Lam=0$ boundary of the far-branch search.

Step (2) bounds the momenta $(\Lam,C_z)$ before the ray search. At each tilt $C_z$, Eq.~\eqref{eq:adm} limits $\Lam$ to $0\le\Lam<\Lmax(C_z)$. The medium that sets this limit is $k^\ast$, and $r_c$ in Eq.~\eqref{eq:rc} indicates whether the target circle or an earlier interface is limiting.
The second bound is set by the target's axial offset. Let's first define $C_{\max}$ that satisfies
\begin{equation}
    Z_0(C_{\max})=\abs{\Delta z},
\end{equation}
where $Z_0$ is from Eq.~\eqref{eq:z0}.
Since $\abs{Z}\ge Z_0(\abs{C_z})$ and $Z_0$ increases with tilt, any ray with $\abs{C_z}>C_{\max}$ has $\abs{Z}>\abs{\Delta z}$, which will miss the target axially.
We therefore search only $\abs{C_z}\le C_{\max}$.

Step (2) also selects the azimuthal sweeps to test. A ray may reach the same transverse point through $\Delta\varphi$, in the opposite sense through $2\pi-\Delta\varphi$, or after an additional full turn through $\Delta\varphi+2\pi$, $4\pi-\Delta\varphi$, and so on.
The list is finite because the sweep on either branch is less than $B=\pi+\sum_{k=1}^{N-1}\arccos(R_k/R_{k-1})$ (Supplement~1, Section~S5); only targets below $B$ need to be considered.
To further shorten the list, we can apply the wrap test: if $C_{\max}<C_\pi$, i.e. $\abs{\Delta z}<Z_0(C_\pi)$ (Eq.~\eqref{eq:wraptest}), no ray within the allowed tilt range sweeps beyond $\pi$, so only $\Delta\varphi$ remains.
Otherwise, step (3) will check the additional targets against both the angular and axial offsets.

\begin{figure}[htbp]
\centering
\paperfig[\linewidth]{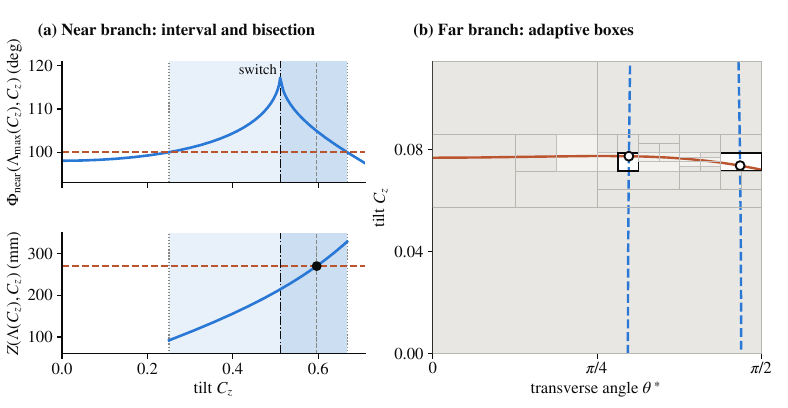}
\caption{Classification in step (3) for the reference cylindrical window. (a) Near branch, $r_X=70$\,mm, $\Delta\varphi=100^\circ$, $\Delta z=270$\,mm: the upper solid curve gives the grazing-limit sweep $\Phinear(\Lmax(C_z),C_z)$; the lower gives $Z(\Lam(C_z),C_z)$ after matching the target sweep. Dashed horizontal lines mark the target offsets. The dash-dotted switch tilt separates two shaded feasible portions, which merge into one interval; the filled circle marks the ray. (b) Far branch, $r_X=86$\,mm, $\Delta\varphi=138^\circ$, $\Delta z=30$\,mm: gray boxes fail the bounds or Krawczyk test. The dashed angular and solid axial level curves intersect at two rays (open circles); their small boxes supply candidates for refinement.}
\label{fig:classification}
\end{figure}

Step (3) searches for rays at each angular target selected in step (2). On the near branch, we first find the tilts that can attain that sweep. In Fig.~\ref{fig:classification}(a), the upper plot shows $\Phinear(\Lmax(C_z),C_z)$, the largest near-branch sweep at each tilt, where the horizontal dashed line marks the target $\Delta\varphi$.
The azimuth equation can be satisfied only when the solid curve meets or exceeds the dashed line, yielding the shaded tilt interval.
To construct that interval, we first cut the full tilt range $[0,C_{\max}]$ at the switch tilts of Eq.~\eqref{eq:cuts}. As shown in Fig.~\ref{fig:classification}(a), the dash-dotted vertical line marks the single cut in this example.
On each side of the cut, the endpoint values bound the total angular reach, as proved in Supplement~1, Section~S3. Then we remove a subinterval if its upper bound is below $\Delta\varphi$, or keep it if its lower bound is at least $\Delta\varphi$, or repeatedly halve it until finding the correct interval.
Finally, adjacent kept subintervals are merged, and the two shaded regions in Fig.~\ref{fig:classification}(a) therefore become one feasible interval across the switch.
At each interior tilt in an interval, $\Phinear(\Lam,C_z)=\Delta\varphi$ determines a unique $\Lam(C_z)$. Then its corresponding axial offset is shown in the bottom plot of Fig.~\ref{fig:classification}(a), where the dashed line marks the target $\Delta z$. Since this offset increases with $C_z$ based on Eq.~\eqref{eq:zetaprime}, the bisection finally finds the solution marked by the filled circle.

On the far branch, a fold can give several $\Lam$ values at the same tilt and angular target, so there is no single $\Lam(C_z)$ curve to follow. The axial offset along a far-branch angular level curve can also turn back with $C_z$ (Supplement~1, Section~S3), so the near-branch bisection no longer applies.
Step (3) therefore searches both momenta together. We set $\Lam=\Lmax(C_z)\sin\theta^\ast$, which maps the admissible domain between successive switch tilts to a rectangle in $(\theta^\ast,C_z)$ (Supplement~1, Section~S4).
Fig.~\ref{fig:classification}(b) illustrates the two constraints: the dashed curve satisfies $\Phifar=\Delta\varphi$, the solid curve satisfies $\Zfar=\Delta z$, and their intersections (open circles) are the two rays. The search does not trace these curves, but instead subdivides the rectangle into boxes of the form $[\theta_a^\ast,\theta_b^\ast]\times[C_a,C_b]$.
For each box, we bound the sweep and axial contribution from every medium using the box endpoints, then sum these bounds to enclose all values of $\Phifar$ and $\Zfar$ throughout the box (Supplement~1, Section~S4). If the box cannot cover both targets, it is then discarded. 
Krawczyk's interval-Newton test can then exclude further boxes (gray) or isolate one ray in a black-outlined box~\cite{krawczyk1969newton,moore2009introduction}.
If the bounds and Krawczyk test leave a box unresolved, we subdivide it and repeat the checks. Any box still unresolved at the numerical resolution limit is kept for refinement in step (4).

Step (4) refines each candidate by Newton's method for Eq.~\eqref{eq:system}, using the angular target selected in step (2). Each update is ensured to stay in the admissible near-branch domain (Section~\ref{sec:newton}), or within the far-branch candidate box. The search solves for the magnitude $\Lam\geq 0$, and then assigns its sign according to the ray's azimuthal direction, since $\Lam=nr d_\varphi$ (Eq.~\eqref{eq:invariants}). For nonnegative integers $m$, targets with $\Delta\varphi+2\pi m$ have the same azimuthal direction as the $\Delta\varphi$ path, so $\Lam$ will keep positive. But targets with $2\pi(m+1)-\Delta\varphi$ have the opposite direction, leading to a negative sign on $\Lam$.
Step (5) uses the calculated $(\Lam,C_z)$ to reconstruct the direction at the pinhole through Eq.~\eqref{eq:pinhole_direction}, and maps the ray to a pixel. 
As a result, the steps return a list of images, with recorded branch, signed momenta and pixel coordinates.

\FloatBarrier

\section{Spherical windows and axial rings}
\label{sec:sphere}

Section~\ref{sec:algorithm} summarizes our forward projection algorithm for the cylindrical windows. In this section, we will discuss about how to generalize the algorithm to the spherical windows.
For a spherical window, symmetry fixes the ray plane through the centre $O$, pinhole $\va*{P}_0$ and target $\va*{X}$ (Section~\ref{sec:invariants}).
In that plane, the interfaces are simple concentric circles, so the cylindrical equations can be simplified to
\begin{equation}
\label{eq:sphere}
    C_z = 0 , \qquad \nu_k = n_k , \qquad b_k = \frac{\Lam}{n_k} .
\end{equation}
Then, the forward projection reduces to solving for the scalar equation
\begin{equation}
    \Phi(\Lam)=t,
\end{equation}
where $t$ is the selected angular targets based on $\Delta\varphi=\angle(\va*{P}_0,O,\va*{X})$. The selection algorithm is shown in Section~\ref{sec:algorithm}.
A spherical window therefore retains the cylinder's near and far branches, bounded angular reach and shadow criterion. The near-branch root is unique by the argument of Section~\ref{sec:convex} with $C_z=0$, and Newton refinement still requires a clamp near grazing incidence.

However, the ray plane is no longer unique on the symmetry axis: a nonradial ray can be rotated about $\va*{P}_0O$ without changing its target. For a target beyond the centre ($\Delta\varphi=\pi$), $\Phifar(0)=\pi$ gives the central ray. If $\Phifar(\Lam)=\pi$ also has a root $\Lam_2>0$, that root describes a ray in every plane through the axis. These rays together form a cone of directions, which appears as a ring around the central image when the camera faces the centre~\cite{nye1999natural}.
For a single sphere ($N=2$), Section~\ref{sec:angularreach} and Supplement~1, Section~S7 give the condition for this second root:
\begin{equation}
\label{eq:ring}
    r_X>r_c , \qquad \Phifar'(0)<0 \qquad\text{and}\qquad \Phifar(\Lmax)>\pi .
\end{equation}
As shown in Supplement~1, Section~S7, a single sphere can produce at most one ring, whereas a multilayer spherical shell can produce more than one.

\begin{figure}[htbp]
\centering
\paperfig[\linewidth]{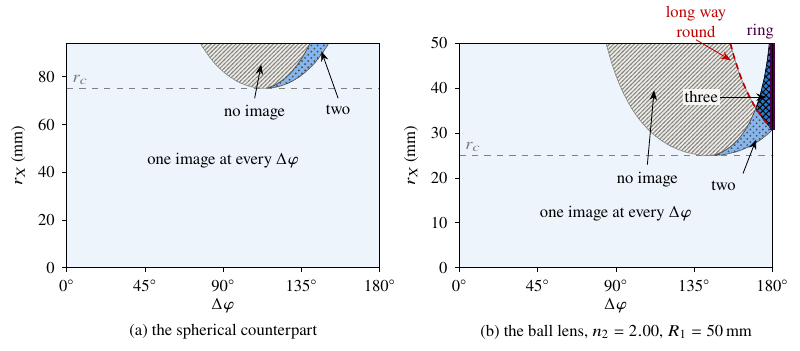}
\caption{Image count inside spherical windows versus $\Delta\varphi=\angle(\va*{P}_0,O,\va*{X})$ and $r_X$. Rotating either map about $\va*{P}_0O$ gives the three-dimensional field. The labeled regions have zero, two or three images; the remainder has one. The dashed boundary marks points reached the long way round, and the thick segment at $\Delta\varphi=180^\circ$ marks axial points with a ring. (a) Spherical counterpart of the reference window. (b) Glass ball with $n_2=2.00$, $R_1=50$\,mm and pinhole 500\,mm from the centre. Counts use all angular targets reached by $\Phinear$ and $\Phifar$ on a $1440\times470$ grid.}
\label{fig:sphmap}
\end{figure}

After considering the axial rings, we now examine how the number of images varies throughout a spherical window.
For an off-axis target, $O$, $\va*{P}_0$ and $\va*{X}$ define a unique ray plane. By rotational symmetry, all points with the same $(\Delta\varphi,r_X)$ have the same image count, so the two-dimensional map determines the image counts throughout the window.
In Fig.~\ref{fig:sphmap}(a), the spherical version of the reference window has one image at every angle below $r_c$. Beyond $r_c$, a shadow opens next to a two-image band, with the caustic between them.
Fig.~\ref{fig:sphmap}(b) shows the additional effect of a far-branch sweep beyond $\pi$, where long-way-round rays reach part of the far-side shadow and produce a three-image cap. The thick axial segment marks points that also have a ring.

\section{Calibration and validation}
\label{sec:validation}
In this section, we are going to validate our forward projection algorithm for both synthetic and real experiments with cylindrical windows. In Section~\ref{sec:wand}, we will first introduce a wand calibration method, which does not require using a calibration board at different depth. Then, the calibration on simulated wand recordings, and the projection against brute-force ray tracing and iterative projection, are tested in Section~\ref{sec:synthetic}. Finally, we discuss the results from our real experiment in Section~\ref{sec:experiment}.

\subsection{Wand calibration through a cylindrical window}
\label{sec:wand}
In order to calibrate the cameras outside the cylindrical windows, the wand calibration is applied here \cite{theriault2014protocol}.
In this method, we use a wand that has two spherical markers with a fixed distance $L$ apart, moved through the measurement volume while the cameras record their images. The inputs are the marker centers, the measured wand length, and the window's interface radii and refractive indices.
The benefit is that there is no need for a fixed calibration board inside the cylinder, enabling the calibration in complex experimental conditions.

\begin{figure}[htbp]
\centering
\paperfig[\linewidth]{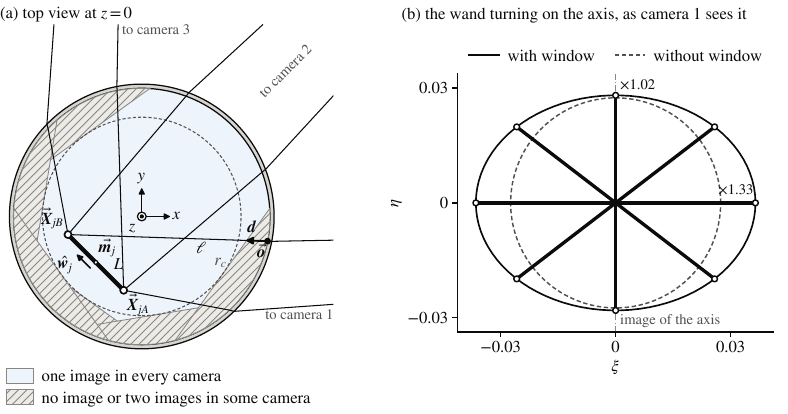}
\caption{Simulated wand geometry for a water-filled acrylic cylinder (radii 100.8 and 97.6\,mm) and three cameras 1.09\,m from its axis at azimuths $0^\circ$, $45^\circ$ and $90^\circ$. (a)~Top view in the pinhole plane $z=0$. The wand is drawn 60\,mm long for legibility; the synthetic calibration uses a 10\,mm wand. Its ends, midpoint, direction and six rays are shown. Camera~1 lies on the $x$ axis by the gauge convention; $\va*{o}$, $\dvec$ and $\ell$ mark the final ray segment from the cylinder to a wand end. Shading gives the image count in this slice only: the light region is single-image for all cameras, and the dashed circle marks $r_c$ at zero tilt. (b)~The normalized image of the drawn wand turning about the axis. Its ends trace an ellipse 1.33 times as wide and 1.02 times as tall as without the cylinder (dashed).}
\label{fig:wandcal}
\end{figure}

\begin{figure}[htbp]
\centering
\paperfig[\linewidth]{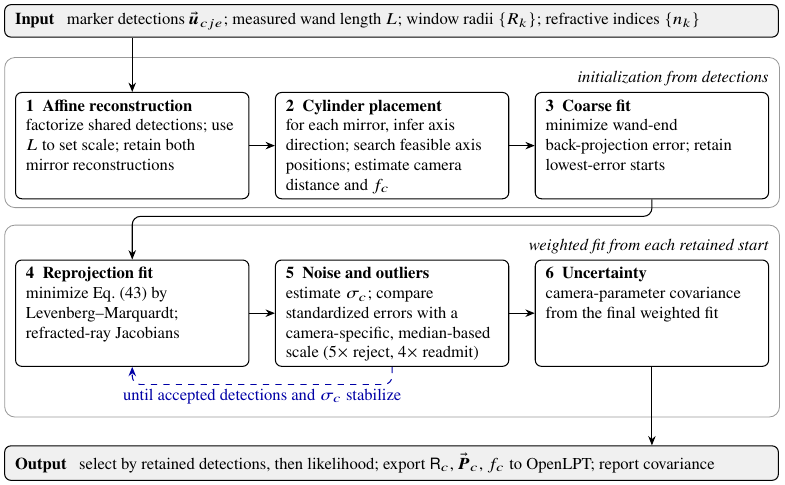}
\caption{The steps of the wand calibration. Steps 1--3 find starting values from the detections alone. Steps 4 and 5 fit the cameras and wand positions, repeating until the observations in use and the estimated noise levels settle; step 6 computes uncertainties for the final fit. The wand remains in the central single-image volume throughout the recording.}
\label{fig:wandpipeline}
\end{figure}

Fig.~\ref{fig:wandcal} illustrates how the wand calibration works.
Panel~(a) shows the known-length wand and the rays to three cameras. The shaded region denotes points with no image or multiple images in the $z=0$ plane.
Therefore, we need to keep the recorded wand inside the central volume without touching the shaded region.
Panel~(b) shows the unequal magnification through the cylindrical window: as the wand turns, its image extends farther along the horizontal direction than along the axial direction.
The overall procedure is then summarized in Fig.~\ref{fig:wandpipeline}. It shows two stages of calibration. In the first stage, steps 1--3 use the detected marker centers to obtain approximate camera and wand positions. In the second stage, steps 4--6 refine these estimates by fitting the predicted images to the detections.

In the first stage, step 1 approximates each camera's projection over the small wand volume by an affine map. A factorization of the marker detections shared by the cameras gives a provisional 3D reconstruction~\cite{hartley2004multiple}.
The measured wand length corrects the possible stretch and gives this provisional reconstruction a physical scale, but leaves two mirror-related reconstructions.
Step 2 estimates the cylinder-axis direction from the direction of least magnification in each view. For each mirror reconstruction, it tries several feasible axis positions relative to the wand, producing several candidate camera configurations.
At each position, the magnifications across and along the axis give initial estimates of each camera's distance from the axis and focal length.
Then, step 3 refines each candidate by minimizing the triangulation errors for the wand ends, then carries the candidates with the smallest coarse-fit errors into the second stage.

\FloatBarrier

In the second stage, each candidate retained after step 3, drawn from either mirror reconstruction, independently undergoes steps 4--6.
We use the cylindrical coordinates as shown in Fig.~\ref{fig:wandcal}(a), with its axis as $z$, and hold the measured window radii and refractive indices fixed.
For each camera $c$, the fit estimates its orientation $\mathsf{R}_c$, pinhole $\va*{P}_c$ and focal length $f_c$, while keeping the principal point at the sensor center.
For each wand position $j$, the wand locations are determined through a midpoint $\va*{m}_j$ and unit direction $\vu*{w}_j$, with $\va*{X}_{jA}=\va*{m}_j-(L/2)\,\vu*{w}_j$ and $\va*{X}_{jB}=\va*{m}_j+(L/2)\,\vu*{w}_j$.
We also set the pinhole of camera 1 at $y=z=0$ on the positive $x$ axis, to avoid multiple solutions from the symmetry.
Then, the physical scale will be set by the known window radii, and the wand length will also be fitted and compared with its measured value as a check of the window parameters (Supplement~1, Section~S8).
With the geometry and unknowns defined, steps 4--6 fit the predicted marker images to the detections, update the noise levels and accepted observations, and estimate parameter uncertainty.

Step 4 adjusts these parameters to minimize the weighted reprojection error,
\begin{equation}
\label{eq:wandcost}
    \sum_{(c,j,e)\in\mathcal{A}}\frac{\norm{\va*{u}_{cje}-\va*{\pi}_c(\va*{X}_{je})}^2}{\sigma_c^2} ,
\end{equation}
where $\va*{u}_{cje}$ is camera $c$'s detection of end $e\in\{A,B\}$ in wand position $j$, $\va*{\pi}_c(\va*{X})$ is the projected image location based on Section~\ref{sec:algorithm}, $\sigma_c$ is that camera's detection-noise level, and $\mathcal{A}$ is the current set of observations.
With a fixed $\mathcal{A}$ and known Gaussian noise levels, this objective gives the maximum-likelihood estimate, and it is optimized with the Levenberg--Marquardt algorithm (Supplement~1, Section~S10).

After fitting in step 4, step 5 estimates each camera's noise level from its residuals and checks for outliers.
When deciding whether a projection error is unexpectedly large, we account for camera noise and how accurately the detections determine that wand position (Supplement~1, Section~S10).
For each camera, step 5 estimates a typical noise-adjusted error from the median of its accepted detections (Supplement~1, Section~S10). Detections exceeding five times this camera-specific value are removed. To prevent the accepted set from oscillating, a removed detection returns only if its error later falls below four times the same value.
Steps 4 and 5 repeat until the accepted detections and noise levels stabilize.

Step 6 estimates each camera parameter's uncertainty from the final weighted fit, accounting for uncertainty in the fitted wand positions (Supplement~1, Section~S10), which is used for the final selection of the fits from the two mirror candidates.
In the output stage, we select the fit that retains the most detections, which is calculated by the likelihood of its residuals under the estimated noise levels. This prevents a fit that has high accuracy by simply ignoring many detections. 
Finally, the selected cameras are saved as OpenLPT camera files.

\subsection{Synthetic validation}
\label{sec:synthetic}

In this part, we focus on synthetic tests against a known truth, where all tests are implemented with OpenLPT, an open-source particle-tracking framework~\cite{tan2020openlpt,tan20233d}. 
First, we test the accuracy and correctness of the forward projection, i.e., whether it returns accurate rays and the correct number of images, in Section~\ref{sec:synthproj}. 
Then, we check whether the wand calibration of Section~\ref{sec:wand} can recover the camera parameters (Section~\ref{sec:wandsynth}). 
Note that the projection tests use the reference window of Section~\ref{sec:setup}, while the calibration tests use the cylinder of the physical experiment.

\subsubsection{Projection accuracy and visibility}
\label{sec:synthproj}

The forward projection is evaluated by two criteria. The first criterion is accuracy: how close the reconstructed line of sight is to the given 3D point. 
To test that, a given 3D point is projected onto the camera, and then the line of sight from each image point is traced through the window by Snell's law, Eq.~\eqref{eq:snell_vec}. Because the image point fixes the incident ray, this trace needs no search. The distance between the traced line of sight and the 3D point measures the accuracy, and a ray counts as correct when this distance is below $10^{-9}$\,m.
The second criterion is correctness: whether the projection returns every image of the point, including an empty list where no ray reaches the point. 
For this, the images of every test point were found by a search that uses only the ray tracer: rays launched from the pinhole on a grid of directions were traced through the window and adjusted by Newton's method until they passed through the point (Supplement~1, Section~S8). 
Each method's answer is then scored as every image, one of two images, no answer, or a false image.

We compared our forward projection algorithm on both criteria with two published iterative projection methods, where the second one is solved in two ways:
\begin{itemize}
    \item Alternating forward ray tracing (AFRT)~\cite{belden2013calibration}. This method iterates on the point where the ray enters the window. From a guess of this entry point, it traces one ray from the pinhole through the entry point into the water, and a second ray backward from the 3-D point, antiparallel to the first in the water, out to the outer surface. The entry point then moves to the midpoint of the two rays' crossings of the outer surface, until the two rays cross both surfaces within $10^{-9}$\,m of each other.
    \item Dewarping method~\cite{paolillo2022perspective}. Unlike AFRT, this method moves a dewarped point $A$ within a fixed plane, which is through the 3D point and normal to the line from the axis to the pinhole. A ray from the pinhole toward $A$, refracted at every interface it crosses, meets this plane at some point. $A$ is updated until the crossing point is the 3D point. Three equations are derived for the two coordinates of $A$, which is optimized by Gauss--Newton or Levenberg--Marquardt method. Paolillo and Astarita found that Gauss--Newton could converge faster, but recommended Levenberg--Marquardt for robustness. Here, we compare both ways, denoted as GN and LM dewarping.
\end{itemize}
In the original papers, each of these methods starts from one trial ray and refines it iteratively, so a run returns at most one image.
We test them first for the accuracy and correctness. In the end, we also try to restart those methods multiple times in the admissible region based on our analysis in Section~\ref{sec:admissible}, to compare the efficiency to find the complete solutions.

\begin{figure}[htbp]
\centering
\paperfig[\linewidth]{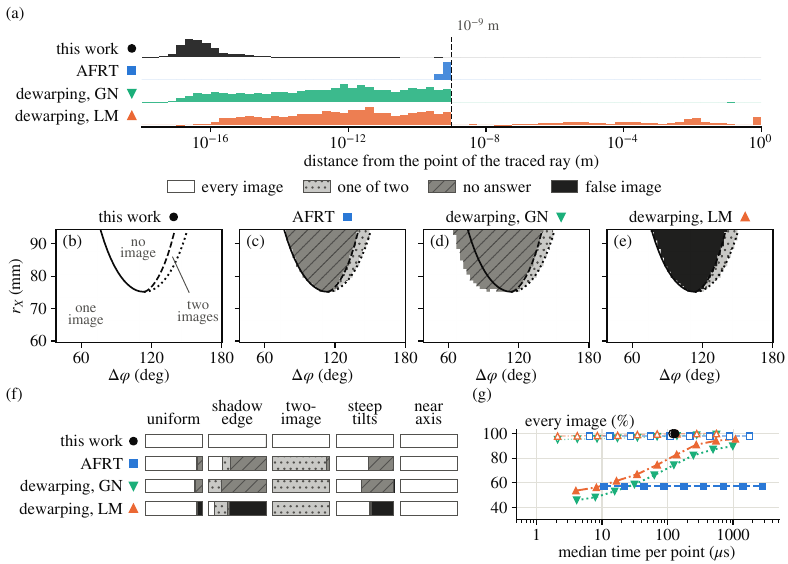}
\caption{Forward-projection accuracy and correctness in the reference cylindrical window. Method symbols: circle, our search; square, alternating forward ray tracing (AFRT); downward/upward triangles, Gauss--Newton/Levenberg--Marquardt dewarping. (a) Single-start ray-to-target distance across all five point sets, with each histogram scaled to its peak. The vertical dashed line marks the $10^{-9}$\,m acceptance threshold; the right edge denotes rays that do not reach the water. (b)--(e) Single-start outcomes on a $40\times50$ grid at $z=0$, plotted against target azimuth $\Delta\varphi$ and radius $r_X$. White denotes a complete image list (including a correct empty list at an unseen point for our search), dots one of two images, diagonal hatching no answer, and black a false image. Solid, dashed, and dotted curves mark inbound grazing, the caustic, and outbound grazing, respectively. The first two bound the shadow; the last two bound the two-image band. (f) Outcome fractions for uniform, shadow-edge, two-image, steep-tilt, and near-axis sets. (g) Fraction of image recovery versus median time per point. Iterative methods use 1--256 starts with independent trace validation; our search uses one call. Open symbols denote the uniform set, and filled symbols denote fraction from all difficult sets.}
\label{fig:baselines}
\end{figure}

The test points include 2000 points sampled uniformly in the water, and four additional sets of 500 difficult points: next to the edges of the shadow, in the two-image band, at steep ray tilts and near the axis. 
To show where each method fails, we also map their results on the plane $z=0$ as shown in Fig.~\ref{fig:baselines}(b)--(e).
All methods are implemented in C++ and run on one thread of Intel Core i9-13900K cpu, for fair comparison of efficiency (Supplement~1, Section~S8).

The results on accuracy is shown in Fig.~\ref{fig:baselines}(a), which shows the error distribution for each method.
Our method will return the most accurate solutions by default, so it gives the lowest error near machine precision.
But the iterative methods stop at the given threshold, so AFRT and dewarping GN have highest error as $10^{-9}$\,m.
However, dewarping LM method sometimes even stops at a higher error, which might be due to its slow convergence rate or vanishing update step.
In summary, both our method and the previous iterative methods can return rays that pass the accuracy test. 
But a small error does not mean a method can find every image or correctly identified a point in shadow, which will be tested in the next criterion.

The correctness test, in Fig.~\ref{fig:baselines}(b)--(f), separates the methods in the two-image and shadow regions, as already mentioned by Paolillo and Astarita~\cite{paolillo2022perspective}. 
A single start returns at most one ray, so none of the iterative methods recovered both images in the two-image set. 
At unseen points, dewarping LM can report a ray that misses the point. AFRT and GN dewarping usually return no ray there, but also return no ray at some visible points. 
An empty answer from these methods therefore cannot truly determine whether a point lies in shadow. 
Our search returned the correct image list at 3999 of the 4000 test points, including the empty list at every unseen point. Its only miss was a ray extremely close to grazing, where rounding affects image recovery, as examined below.

One way to improve the iterative methods is to restart each one from many rays and keep only the rays that tracing confirms.
We restart each iterative method from up to 256 rays per point: the straight line first, then rays drawn uniformly over the admissible domain of $(\Lam,C_z)$ derived in Section~\ref{sec:admissible}. A ray is kept when its trace reaches the point within $10^{-9}$\,m, so a restarted method can return several images and never a false one.
Multiple restarts will increase the total computation cost, and the benefits it brings also depends on the point is, as shown in Fig.~\ref{fig:baselines}(g).
For points sampled uniformly in the water, most of which have a single image, a few restarts work well. The open symbols show that the iterative methods can find almost every image in less time than our method.
However, for points in the difficult sets, the iterative methods cannot work well even with 256 restarts.
With more restarts, both dewarping methods graduately increase their correctness, while AFRT cannot improve at all. For these difficult points, the iterative methods cost much more than our method and still cannot guarantee to be fully correct. 
These failures matter in dense particle tracking~\cite{schanz2016shake,tan2020openlpt}, where a missed or false image can lose or identify a wrong particle, and multiple restarts increase the cost of each projection.
In short, restarts can make iterative projection complete at most points, but it cannot guarantee the complete image list.

Compared to the iterative methods, our method can instead find the complete image list in a single call, as shown in Section~\ref{sec:algorithm}.
It can only lose a ray near the grazing limit due to the machine precision, where $1-\Lam/\Lmax$ is below about $10^{-13}$.
Furthermore, the cost of our search is nearly the same at the difficult points as elsewhere, while the iterative methods with multiple restarts take an order of magnitude longer and still miss images.
Therefore, our method provides accuracy and full correctness for calibration and particle tracking at a predictable cost.

\subsubsection{Wand calibration}
\label{sec:wandsynth}
We test the calibration of Section~\ref{sec:wand} on synthetic wand recordings, and compare with the true camera parameters.
To match the experiment of Section~\ref{sec:experiment}, the recordings are made through the water-filled acrylic cylinder with the same size.
We simulate 500 frames of a 10\,mm wand moving within the single-image domain using OpenLPT's projection, and then add Gaussian noise.

Table~\ref{tab:wand} summarizes calibration accuracy on the three-camera setup of Fig.~\ref{fig:wandcal}(a). With 0.1\,px of detection noise, the focal-length error and the wand length error are below 0.1\%. At points not used for calibration, the projection error is below 0.01\,px, and the triangulated 3D shape agrees with the truth to below 1\,$\mu$m.

\begin{figure}[htbp]
\centering
\paperfig[0.55\linewidth]{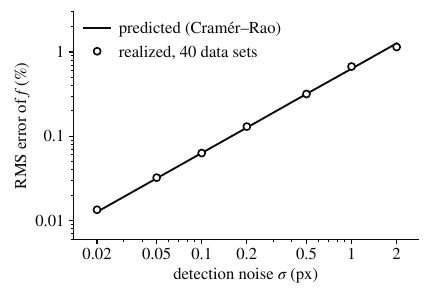}
\caption{Accuracy of the synthetic wand calibration through the cylinder of the experiment: root-mean-square error of the focal length against the detection noise on the reference rig of Fig.~\ref{fig:wandcal}, realized over 40 data sets per noise level (circles) and predicted by the Cram\'er--Rao bound (line), each the root mean square over the three cameras.}
\label{fig:wandsynth}
\end{figure}

The precision limit of the calibration can be estimated from the image noise and camera location based on the Cram\'er--Rao bound.
As shown in Table~\ref{tab:wand}, the calibration error match very well with the predicted one. We also plot how the focal-length error grows with detection noise in Fig.~\ref{fig:wandsynth}, and it also matches with the predicted error. 
Thus, the calibration method will approache to the precision limit and provides useful error bars.

\begin{table}[htbp]
\caption{Calibration errors on the reference rig at 0.1\,px of detection noise}
\label{tab:wand}
\centering
\footnotesize
\begin{tabular}{@{}lrr@{}}
\toprule
quantity & realized & predicted \\
\midrule
focal length $f$ error & 0.063\% & 0.063\% \\
wand length error & 0.024\% & 0.023\% \\
projection error on unseen points & 0.0079\,px & 0.0077\,px \\
3D shape error on unseen points & 0.76\,$\mu$m & -- \\
\bottomrule
\end{tabular}
\\[1mm]\footnotesize Realized: root-mean-square error over 40 noisy data sets for $f$ and the 3D error, 20 for projection error, and two data sets in five-fold holdout for wand length. Predicted: the Cram\'er--Rao bound at the true parameters; --: not computed. Unseen points: a $7\times7\times7$ grid over the wand's $\pm20$\,mm volume. Projection error is the reprojection residual after triangulating their noise-free images with the fitted cameras. For the 3D shape error, the entire reconstructed grid is shifted and rotated to match the true grid before computing the root-mean-square point error.
\end{table}

The above analysis shows that the calibration works well in the synthetic tests, and we will show a real experiment on camera calibration and particle tracking with two camera in the following.

\subsection{Experiment}
\label{sec:experiment}
In this part, we apply the wand calibration of Section~\ref{sec:wand} and the forward projection algorithm in Section~\ref{sec:algorithm} to a real experiment. As mentioned in Section~\ref{sec:wandsynth}, we use a water-filled acrylic cylinder with an outer diameter of 201.6\,mm and a 3.175\,mm wall.
Three high-speed cameras (two Phantom VEO~710L with $1280\times800$ pixels of 20\,$\mu$m and a Phantom v2640 with $2048\times1952$ pixels of 13.5\,$\mu$m, all with 180\,mm macro lenses) view the water through the wall with backlighting. Camera 2 is $42^\circ$ from camera 1 around the axis, and camera 3 sits almost directly below camera 1, which is about 0.3\,m higher than the other two and looks down at $23^\circ$.
In the experiment, we first perform wand calibration, and then apply Lagrangian particle tracking with OpenLPT for a swirl.

\begin{figure}[htbp]
\centering
\paperfig[\linewidth]{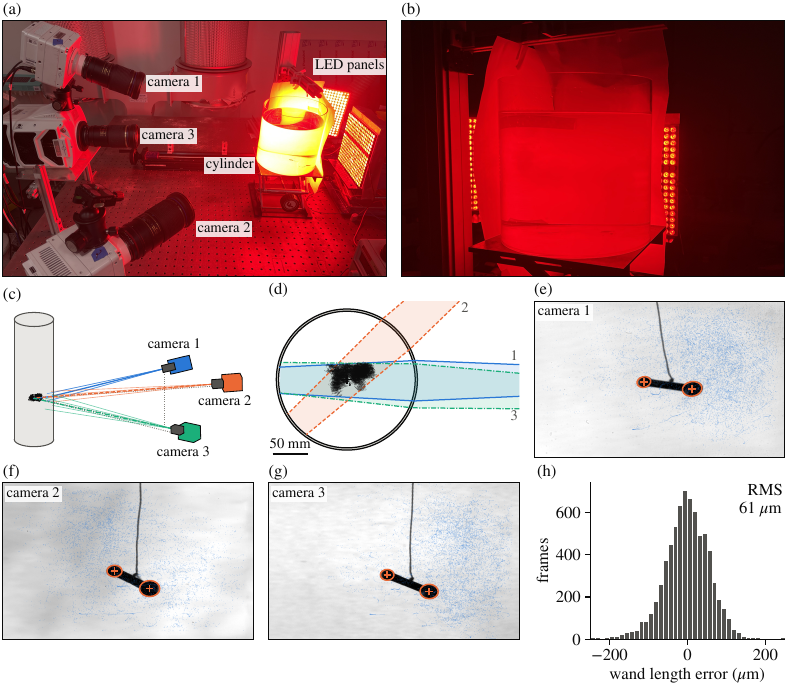}
\caption{The experimental setup and the wand calibration of its three cameras through the water-filled cylinder. (a)--(b)~Images of real experimental setup. (c)~The calibrated rig to scale, in an oblique view: the cylinder (gray), the cameras, the lines of sight through the corners of each camera's image to the cylinder and through its center to the wand, and the wand-end positions of the calibration (dots); dotted lines run from the axis to each camera's foot on the plane of the wand's mean height and up to the camera. (d)~The rig seen along the cylinder's axis: the wall (gray), the field of view of each camera in its central image row, traced through the wall (shaded; edges solid for camera 1, dashed for camera 2, dash-dotted for camera 3), and the wand-end positions (dots). (e)--(g)~One frame of the wand recording in cameras 1 to 3, with the ellipses fitted to the balls and their centers (orange), and the centers of all ball images that the calibration used (blue dots); camera 3's frame is cropped to 1280 of its 1952 rows. (h)~The wand length triangulated independently in each of 7175 frames of the fit, minus its known 10\,mm.}
\label{fig:expwand}
\end{figure}

Fig.~\ref{fig:expwand}(a) and (b) show the experimental setup: three cameras look at the cylinder from one side with backlit from two LED panels. The calibrated system is illustrated in Fig.~\ref{fig:expwand}(c) and (d). The fitted pinholes are 0.79, 0.93 and 0.87\,m from the axis, and the wall bends their fields of view, which overlap near the axis where the wand moves.
Fig.~\ref{fig:expwand}(e)--(g) show one frame of the wand recording.
In each frame a wand with 3 and 4\,mm balls with 10\,mm central distance moves around in the central single-image domain, which is recorded at 100 Hz for 130\,s.
Through the cylinder, each ball images as an ellipse about 1.2 to 1.3 times wider across the axis than along it, which is the unequal magnification, as illustrated in Fig.~\ref{fig:wandcal}(b).
Thus, the balls are identified by ellipse fitting and only 7212 frames are kept, because both balls need to be identified clearly in at least two cameras. The used centers are marked as blue dots in Fig.~\ref{fig:expwand}(e)--(g).

As shown in Table~\ref{tab:experiment}, the reprojection errors of the three cameras are 0.5 to 0.8 px, which is larger than the synthetic cases due to the noise of object identification. But this is still good enough for 3D particle tracking with subpixel accuracy. The relative error for 3D wand length reconstruction is only 0.61\%, and its distribution is close to a Gaussian centered at zero, as shown in Fig.~\ref{fig:expwand}(h).

\begin{table}[htbp]
\caption{Calibration and tracking quality in the experiment}
\label{tab:experiment}
\centering
\footnotesize
\begin{tabular}{@{}lr@{}}
\toprule
quantity & value \\
\midrule
\multicolumn{2}{@{}l}{\emph{wand calibration, 7212 frames}} \\
reprojection error (camera 1/2/3) & 0.74 / 0.53 / 0.78\,px \\
wand length error & 61\,$\mu$m (0.61\%) \\
\midrule
\multicolumn{2}{@{}l}{\emph{tracking, 250 frames}} \\
particles detected per frame, camera 1 / 2 / 3 & 459 / 959 / 3387 \\
particles tracked per frame & 343 \\
number of tracks (positions) & 425 (69\,846) \\
track length median (90th percentile) & 184 (250) \\
number of long tracks ($\geq100$ frames) & 320 \\
tracked particle reprojection error & 0.27--0.84\,px \\
velocity range & 113--252\,mm/s \\
\bottomrule
\end{tabular}
\end{table}

\begin{figure}[htbp]
\centering
\paperfig[\linewidth]{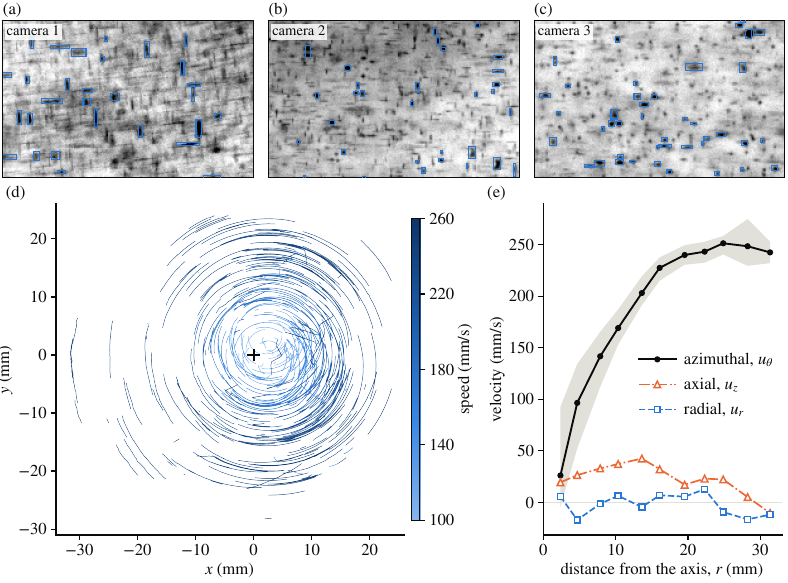}
\caption{Particle tracking through the cylinder with three cameras. (a)--(c)~A $200\times125$\,px crop of one frame in cameras 1 to 3, with the particles dark against the backlight; the boxes outline the particle images identified in the frame. (d)~All 411 tracks long enough to smooth (at least 19 frames), smoothed with a Gaussian filter of width 3 steps, viewed along the cylinder's axis (+) and shaded by speed, in the frame of Fig.~\ref{fig:expwand}(d): $x$ points towards camera 1. (e)~The median azimuthal, axial and radial velocities of the smoothed tracks against the distance from the axis, in 3\,mm bins; the shading spans the interquartile range of the azimuthal velocity.}
\label{fig:exptrack}
\end{figure}

Next, we perform Lagrangian particle tracking with OpenLPT~\cite{schanz2016shake,tan2020openlpt}, for a swirling flow inside the cylinder. 60\,$\mu$m polyamide particles are used as tracer particles, and the video is recorded at 4000 Hz for 0.25\,s, of which we track the first 250 frames.
Since the cylinder window will stretch the dimension across its axis, each particle therefore images as a short horizontal or vertical line rather than a round spot in cameras 1 and 2, as shown in Fig.~\ref{fig:exptrack}(a) and (b). The orientation and length of the pattern depend on the particle's depth along the line of sight. Camera 3, whose images are less blurred, records most particles as compact dots (Fig.~\ref{fig:exptrack}(c)).
We detect the particles as line segments, use their shapes to enhance the stereo match, and require every particle to be matched in all three cameras.

The tracking quality is summarized in Table~\ref{tab:experiment}. In the 250 tracked frames, about 340 particles are tracked in each frame, 48 tracks persist from the beginning till the end, and the total number of tracks is 425 with a median track length of 184 frames. The projection error of the tracked particles is less than 1 px in every camera. Since every particle must be seen by all three cameras, few tracks are ghosts: a control run in which no match can be real builds at most 6\% as many long tracks as the real one (Supplement~1, Section~S8).
The tracks are shown in Fig.~\ref{fig:exptrack}(d) without any selection. The trajectories of particles clearly show a swirling flow pattern as expected. And their velocity statistics are plotted in Fig.~\ref{fig:exptrack}(e). It shows that the azimuthal velocity rises steeply away from the axis and levels off at about 250\,mm/s beyond 20\,mm, while the radial velocity stays close to zero and the axial velocity, a slow drift along the axis, stays below 45\,mm/s, which validates against the real scenario.

In summary, these results show that our forward projection algorithm together with the wand calibration works well through cylinder windows, which can be used for high-quality particle tracking with three cameras.
It provides a robust and easy way to calibrate multi-camera system without the need of calibration target, which is very useful for many experiments or applications that has limited space.

\section{Conclusion}
\label{sec:conclusion}

In this paper, we develop a complete forward-projection algorithm for multilayer coaxial cylindrical or concentric spherical systems. 
First, closed-form expressions are derived to describe the ray travelling throught the curved windows. They are based on the conserved ray momenta $(\Lam, C_z)$, which reduce the complex projection to a search in one or two variables. 
Then, we can divide the projection into the near-branch and far-branch problems, where the near-branch problem can only have one solution at most. 
For the far branch, we also propose an algorithm that can find all possible solutions. 
Following the algorithm in Section~\ref{sec:algorithm}, we are able to find the complete solution with theoretical guarantee, which can hardly be achieved with traditional iterative method \cite{paolillo2022perspective}.

The validation results also show that our forward projection is accurate and robust. 
Our C++ implementation returned the full correct image list, except for one missing ray due to floating-point rounding limit. 
In contrast, none of the single-start iterative methods is able to find both images in the multi-image domain, and multiple restarts can still miss images at difficult points. 
With our robust projection algorithm, the simple wand calibration works well in both synthetic tests and the real experiment, which has only 0.61\% error on wand-length reconstruction. 
It is also successfully tested for three-camera tracking of a swirling flow with OpenLPT \cite{tan2020openlpt,tan20233d}, which sheds light for 3D pipe flow measurements, accurate underwater reconstruction, and many real applications.

\appendix
\renewcommand{\thefigure}{A\arabic{figure}}
\renewcommand{\theHfigure}{A\arabic{figure}}
\setcounter{figure}{0}

\section{Index convention}
\label{app:index}
\begin{figure}[htbp]
\centering
\paperfig[\linewidth]{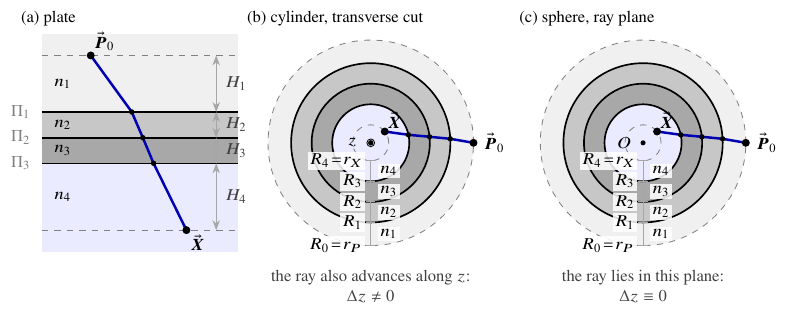}
\caption{Index convention, drawn for the same four-medium stack in all three geometries. (a) Flat plates: medium $i$ has thickness $H_i$ along the normal, and $H_1$ and $H_N$ are measured from the pinhole $\va*{P}_0$ and the object point $\va*{X}$ to the nearest interface. (b) Coaxial cylinders and (c) concentric spheres: medium $k$ lies between the radii $R_k$ and $R_{k-1}$. Solid lines are physical interfaces; dashed lines are not. In (b) and (c) the dashed circles carry $R_0=r_P$ and $R_N=r_X$, which enter the formulas exactly like interface radii but are only where the ray starts and ends. With $N=4$ media there are $N-1=3$ interfaces and $N-2=2$ solid walls, yet five radii $R_0,\dots,R_4$ appear in the sums.}
\label{fig:index}
\end{figure}

Fig.~\ref{fig:index} draws the index convention of Sections~\ref{sec:setup} and~\ref{sec:flat} for the same four-medium stack in all three geometries. In each, medium $1$ holds the pinhole and medium $N$ the object point, and the number of media, not of surfaces, sets the size of the problem.

\begin{backmatter}
\bmsection{Funding}
TBD.

\bmsection{Acknowledgment}
TBD.

\bmsection{Disclosures}
The authors declare no conflicts of interest.

\bmsection{Data availability}
The Python reference implementation, the verification scripts, the scripts that print every number, table and figure of this paper, and the data behind Figs.~\ref{fig:baselines} and~\ref{fig:wandsynth} will be available in a public repository (TBD). The C++ implementation of the projection and of the wand calibration is available at \url{https://github.com/clockj/pyOpenLPT-refraction}, which will be merged into the main repository of OpenLPT~\cite{tan2020openlpt}, \url{https://github.com/JHU-NI-LAB/OpenLPT_GUI}. Supplement~1, Section~S8 names the script behind each table and figure.

\bmsection{Supplemental document}
See Supplement 1 for supporting content.

\end{backmatter}

\bibliography{refract}

\end{document}